# Metallicity of Active Galactic Nuclei from ultraviolet and optical emission lines – I. Carbon abundance dependence

O. L. Dors[1]⋆, C. B. Oliveira[1], M. V. Cardaci[2,3], G. F. Hägele[2,3], I. N. Morais[1], X. Ji[4,5], R. A. Riffel[6], R. Riffel[7], M. Mezcua[8,9], G. C. Almeida[1], P. C. Santos[1], M. S. Z. de Mellos[6]

[1] *Universidade do Vale do Paraíba, Av. Shishima Hifumi, 2911, Zip Code 12244-000, São José dos Campos, SP, Brazil*
[2] *Facultad de Ciencias Astronómicas y Geofísicas, Universidad Nacional de La Plata, Paseo del Bosque s/n, 1900 La Plata, Argentina*
[3] *Instituto de Astrofísica de La Plata (CONICET-UNLP), La Plata, Avenida Centenario (Paseo del Bosque) S/N, B1900FWA, Argentina*
[4] *Kavli Institute for Cosmology, University of Cambridge, Madingley Road, Cambridge CB3 0HA, UK*
[5] *Cavendish Laboratory, University of Cambridge, 19 JJ Thomson Avenue, Cambridge CB3 0HE, UK*
[6] *Departamento de Física, Centro de Ciências Naturais e Exatas, Universidade Federal de Santa Maria, 97105-900, Santa Maria, RS, Brazil*
[7] *Departamento de Astronomia, Universidade Federal do Rio Grande do Sul, Av. Bento Gonçalves 9500, Porto Alegre, RS, Brazil*
[8] *Institute of Space Sciences (ICE, CSIC), Campus UAB, Carrer de Magrans, E-08193 Barcelona, Spain*
[9] *Institut d'Estudis Espacials de Catalunya (IEEC), Carrer Gran Capita, E-08034 Barcelona, Spain*

**ABSTRACT**

Metallicity ($Z$) estimates based on ultraviolet (UV) emission lines from the narrow-line regions (NLRs) of active galactic nuclei (AGNs) have been found to differ from those derived from optical lines. However, the origin of this discrepancy ($ZR$) remains poorly understood. To investigate the source of $ZR$, we compiled from the literature the fluxes of narrow near-UV ($1000 < \lambda(\text{Å}) < 2000$) and optical ($3000 < \lambda(\text{Å}) < 7000$) emission line measurements for a sample of 11 AGNs (9 at $z < 0.4$ and 2 at $z \sim 2.4$). Metallicity values for our sample were derived using a semi-empirical calibration based on the $C43 = \log[(\text{C\,{\sc iv}}\lambda 1549 + \text{C\,{\sc iii}}]\lambda 1909)/\text{He\,{\sc ii}}\lambda 1640]$ emission-line ratio and compared with those obtained via direct measurement of the electron temperature ($T_e$-method) and via calibrations based on optical emission-lines. The source of the discrepancy was investigated in terms of the ionization parameter ($U$), electron density ($N_e$), and carbon abundance (C/H). We found a weak correlation between $ZR$, $U$ and $N_e$. However, a moderate correlation was observed between $ZR$ and direct estimates of C/H, suggesting that the previously assumed (C/O)-$Z$ relations in photoionization models used to derive UV carbon-line calibrations may not be valid for AGNs. By combining a large set of abundance estimates for local star-forming regions with those of our AGN sample, we derived a new (C/O)-$Z$ relation. Comparisons between the results of photoionization models that assume this new abundance relation and the UV observational data of our sample produce $Z$ values derived from the $C43$ index that are consistent with those obtained using the $T_e$-method.

## 1 INTRODUCTION

Recent observations of high-redshift systems by the James Webb Space Telescope (JWST) are providing important information on the metallicity ($Z$) of the gas phase of the earliest galaxies in the Universe. The specific $Z$ pattern of a galaxy depends on the star formation history and chemical enrichment of its Interstellar Medium (ISM). Thus, $Z$ estimates of objects at very high-$z$ uncover how the first stars formed and evolved and their interplay with the ISM in the early epoch of the Universe.

Due to the distance, most of the observed spectra of objects at $z > 6$ are in the ultraviolet (UV) wavelength range [$1000 < \lambda(\text{Å}) < 2000$], where emission-line ratios such as C\,{\sc iii}]$\lambda 1909$/O\,{\sc iii}]$\lambda 1666$ (e.g. Garnett et al. 1995a), N\,{\sc v}$\lambda 1240$/He\,{\sc ii}$\lambda 1640$ (e.g. Hamann & Ferland 1992) and $C43 = (\text{C\,{\sc iii}}]\lambda 1909 + \text{C\,{\sc iv}}\lambda 1549)/\text{He\,{\sc ii}}\lambda 1640$ (Dors et al. 2014, 2019; Pérez-Montero et al. 2023) can be used as a $Z$ tracer (see also Zhu et al. 2024). However, $Z$ estimates through rest-frame UV lines are subject to several caveats:

⋆ E-mail: olidors@univap.br (OLD)

- The unique hydrogen reference line in the UV spectrum is Ly$\alpha$, a resonance line that can be enhanced by the presence of the He II$\lambda$1215.1 and O V]$\lambda$1213.8, $\lambda$1218.3 emission lines (e.g. Shields et al. 1995; Humphrey 2019). This fact results in imprecise dust reddening correction (generally not considered or taken from Balmer decrement) and forces us to use the He II$\lambda$1640 emission line as a reference.

- Photoionization models are generally employed to obtain calibrations between UV line ratios and $Z$. These calibrations are known as strong-line methods. In photoionization models it is required to assume pre-established abundance relations [e.g. (C/O)-(O/H), (N/O)-(O/H)] as input parameters. These abundance relations are barely known for high-$z$ objects (e.g. Cameron et al. 2023; Isobe et al. 2023; Ji et al. 2024; Rizzuti et al. 2024; D'Eugenio et al. 2024), resulting in not so well-established UV emission-line calibrations in comparison to the optical ones (e.g. Zhu et al. 2024).

- Carbon and oxygen are depleted onto dust-grains in the diffuse ISM (e.g. Jenkins 2009) by uncertain factors and estimated mostly just for nearby Star-Forming Regions (SFs: H II regions and H II galaxies; e.g. Garnett et al. 1995a,b; Peimbert & Peimbert 2010). This uncertainty can deeply impact on the derived $Z$ values based on lines emitted by these elements by a factor of up to $\sim 3$ (e.g. see Roman-Duval et al. 2022 and references therein).

- Any proposed method to estimate $Z$ must be compared with direct estimates, i.e. with those via $T_{\rm e}$-method[1], the most reliable method (see e.g. Hägele et al. 2006, 2008; Toribio San Cipriano et al. 2017). Comparison between $Z$ estimations based on UV lines and those through $T_{\rm e}$-method has been carried out mainly for SFs (e.g. Byler et al. 2020).

Byler et al. (2020) found, for SFs located in a wide range of redshift ($0.04 > z \gtrsim 1.5$), that UV-based metallicities poorly correlate with those from optical lines. Additionally, Rigby et al. (2021) found, for the lensed galaxy SDSS J1723+3411 ($z = 1.3293$), that UV-only emission lines either can not be used to estimate the metallicity or dramatically underestimate it. Finally, for the AGN system GS-3073 ($z \sim 5.5$) Ji et al. (2024), by using the $T_{\rm e}$-method, derived log(N/O) abundance ratio values of $\sim -1.1$ dex and $\sim 0.46$ dex when optical and UV emission lines are considered, respectively. In general, $Z$ values for SFs derived through UV lines are overestimated by up to $\sim 0.6$ dex concerning those from optical lines (e.g. Byler et al. 2020; Mingozzi et al. 2022; Llerena et al. 2023). The source of such $Z$ discrepancies has been mainly investigated for SFs and has been attributed to, for instance, the limitation in theoretical ionizing spectrum assumed in photoionization models and the contribution from stellar wind emission to the ISM enrichment (Byler et al. 2020).

Opposite to SFs, for Active Galactic Nuclei (AGNs) a comparison between UV-based metallicities from strong-line methods and those from the $T_{\rm e}$-method[2] is barely found in the literature. This is partially due to the lack of spectroscopic data, for the same sample of objects, taken in both ultraviolet and optical wavelength ranges (see Table 2 of Dors et al. 2019). As noted in previous studies, the UV and optical-based metallicities of AGNs, such as for SFs, seem to not conciliate with each other. For instance, no clear correlation between the metallicity of local universe AGNs ($z < 0.4$) and the host galaxy stellar mass ($M_\star$) is obtained when optical lines are employed in the $Z$ derivation (e.g. Dors et al. 2020a; Peluso et al. 2023; Li et al. 2024). In opposite, UV-based metallicities of AGNs at high-$z$ indicate a clear positive correlation with the $M_\star$ (e.g. Matsuoka et al. 2018; Dors et al. 2019). Although this disagreement can be due to the effects of completeness, redshift bias and/or aperture effects of the AGN sample (e.g. Thomas et al. 2018a; Carr et al. 2023; Armah et al. 2023, 2024), part of it can be attributed to the methods employed in the $Z$ estimation. For instance, calibrations between UV emission lines and $Z$ have been obtained by using photoionization models (e.g. Nagao et al. 2006; Dors et al. 2014, 2019; Pérez-Montero et al. 2023; Zhu et al. 2024). In general, these models suffer limitations in the sense that the assumed abundance relations [e.g. (N/O)-(O/H), (C/O)-(O/H)], derived for local SFs and used as input parameters in photoionization models (e.g. Feltre et al. 2016), could deviate from those for AGNs (e.g. Baldwin et al. 2003; Dors et al. 2019; Ji & Yan 2020).

Pérez-Montero et al. (2023), by using the HII-CHI-MISTRY-UV code, which is based on a comparison between photoionization model results and observational spectroscopic data, found C/O abundances in local AGNs higher (by $\sim 0.3$ dex) than those for SFs with similar $Z$. Moreover, Nakajima & Maiolino (2022) pointed out that although the C/O and N/O abundance ratios are well established for H II regions, they are subject to significant dispersion, possibly caused by several processes including variations in star formation efficiency (e.g. Mollá et al. 2006) or gas exchange between galaxies and the surrounding intergalactic medium (e.g. Edmunds 1990; Köppen & Hensler 2005). This dispersion could also be increased by some systematic uncertainties at high redshift (see also Gutkin et al. 2016) and it is unknown for AGNs.

Additional discrepancy is also noted when UV and optical emission lines are used in the ionization parameter[3] ($U$) derivation. In fact, whilst $U$ estimates through narrow optical emission lines of AGNs indicate log$U$ ranging from $\sim -4.0$ to $\sim -2.5$ dex (e.g. Castro et al. 2017; Carvalho et al. 2020; Carr et al. 2023), UV lines derivations result in higher ($-2.0 \lesssim \log U \lesssim -0.5$) $U$ values (e.g. Nagao et al. 2006; Feltre et al. 2016; Dors et al. 2019; Pérez-Montero et al. 2023; Mingozzi et al. 2024). As in the case of the metallicity, the source of discrepancy in the ionization parameter derived from UV and optical lines remains unclear. $U$ estimates significantly impact the derivation of $Z$, as emission line ratios sensitive to $Z$ can also depend on this nebular parameter (e.g. McGaugh 1991).

To investigate the AGN metallicity discrepancy derived when UV and optical emission lines are employed in the derivation of this parameter, in the present study, we collected optical and UV emission-line fluxes of eleven objects from the literature and we derived their $Z$. We assumed as reference $Z$ values derived through the $T_{\rm e}$-method and from the empirical calibration relied on it proposed by Dors (2021). The $Z$ discrepancy is investigated regarding its dependence on the electron density, ionization parameter and carbon abundance. The paper is organized as follows. In Section 2 the methodology (observational data, metallicity and elemental abundance estimates) are described. The results and discussion are presented in Sect. 3, while the conclusions in Sect. 4.

## 2 METHODOLOGY

To compare the AGNs gas-phase metallicities derived from UV and optical emission lines, the following methodology is assumed:

---

[1] For a review of the $T_{\rm e}$-method see Pérez-Montero (2017) and Peimbert et al. (2017).
[2] For a comparison between oxygen abundance estimates of AGNs based on distinct optical emission lines see Dors et al. (2020a).

[3] The ionization parameter, $U$, is defined as the ratio of the flux of hydrogen-ionizing photons to the total hydrogen density.

(i) For each object, fluxes of observational emission lines in the UV [1000 < $\lambda$(Å) < 2000] and optical [3000 < $\lambda$(Å) < 7000] wavelength ranges were compiled from the literature.

(ii) Observational line intensity ratios were used to estimate metallicities based on the $T_e$-method and strong-line methods.

(iii) Direct abundance estimates, nebular parameters and photoionization models were employed to investigate the source of the metallicity discrepancy (if it exists) between estimates derived from UV and optical emission line ratios.

The following subsections describe each step of the procedures outlined above.

### 2.1 Observational data

We compiled UV and optical emission lines from the literature for objects classified as AGNs. Their classifications were carried out by the authors of the original works from which the data were collected either by considering the presence of coronal emission lines emitted by elements with high ionization degree (e.g. [Ne v]$\lambda$3426, [Fe x]$\lambda$6374) or by using emission line diagnostic diagrams (e.g. [O iii]$\lambda$5007/H$\beta$ versus [N ii]$\lambda$6584/H$\alpha$; Baldwin et al. 1981). We adopt the following criteria in the AGNs selection.

- We only selected AGNs whose optical [O ii]$\lambda$3726 + $\lambda$3729 (hereafter [O ii]$\lambda$3727), H$\beta$, [O iii]$\lambda$5007, H$\alpha$, [N ii]$\lambda$6584, [S ii]$\lambda$6716, $\lambda$6731 and rest-frame UV C iii]$\lambda$1909, C iv$\lambda$1549, He ii$\lambda$1640 emission lines were measured. These emission lines allow us to estimate metallicities through strong-line methods (e.g. Storchi-Bergmann et al. 1998; Nagao et al. 2006; Dors et al. 2020a).

- Although not used as a selection criterion, the fluxes of the optical [O iii]$\lambda$4363 and [N ii]$\lambda$5755 auroral lines, along with those of the Ly$\alpha$$\lambda$1216, O iii]$\lambda$1664 and C ii]$\lambda$2325 lines were compiled when available.

- Additionally, when available, the fluxes of the He ii$\lambda$4686 and He i$\lambda$5876 emission lines were compiled from the literature. These lines are necessary to estimate the ionization correction factor (ICF) of the oxygen (e.g. Flury & Moran 2020).

- We consider only the narrow component of the lines above, i.e. H$\alpha$ or Ly$\alpha$ with full-width half maximum (FHWM) lower than 1000 km s$^{-1}$. This criterion does select objects for which only low-velocity shocks (i.e. cloud-cloud collisions) can be present, hence gas dominated by highly turbulent material is not expected (e.g. Daltabuit & Cox 1972; Daltabuit et al. 1978; Dopita & Sutherland 1996; Bottorff et al. 2000; Foschini 2002; Mazzolari et al. 2024).

We found 11 AGNs that fulfill all the criteria above, being 9 objects in the local universe ($z < 0.04$) and 2 objects at high redshift ($z \sim 2.4$). In Table 1 the selected AGNs together with the references from which the optical and UV spectroscopic data were compiled, their redshifts, AGN classification, and X-ray luminosities (when available) are listed.

Even though the selected objects were already classified as AGNs, we perform an additional classification relied on the He ii$\lambda$4686/H$\beta$ versus [N ii]$\lambda$6584/H$\alpha$ diagram proposed by Shirazi & Brinchmann (2012). According to Molina et al. (2021), objects with

$$\log(\text{He II}\lambda 4686/\text{H}\beta) \gtrsim -1.0 \quad (1)$$

are classified as AGNs and, otherwise, as SFs (see also Scholtz et al. 2023). In principle, this diagnostic diagram is more reliable than the classical [O iii]$\lambda$5007/H$\beta$ versus [N ii]$\lambda$6584/H$\alpha$ (Baldwin et al. 1981) because the latter tends to misclassify as SFs those AGNs with

$Z \lesssim 0.5 \, Z_\odot$ (e.g. Groves et al. 2006; Feltre et al. 2016; Hirschmann et al. 2023; Dors et al. 2024a; Mazzolari et al. 2024) that, in general, are located in galaxies with low mass ($M_\star \lesssim 10^{9.5} \, \text{M}_\odot$, e.g. Izotov & Thuan 2008; Bykov et al. 2024; Mezcua & Domínguez Sánchez 2024). In Figure 1, bottom left panel, the He ii/H$\beta$ vs. [N ii]/H$\alpha$ diagram containing the observational data of our sample is shown. In this diagram all objects occupy the AGNs region while in the [O iii]/H$\beta$ vs. [N ii]/H$\alpha$ diagnostic diagram (top left panel of Fig. 1) one object of our sample, III Zw77, is located in the zone occupied by SFs, below the maximum starburst lines proposed by Kauffmann et al. (2003) and Kewley et al. (2006). III Zw77 presents a very low $N2 = \log([\text{N II}]\lambda 6584/\text{H}\alpha)$ line intensity ratio that, according to the $Z$-$N2$ calibration proposed by Carvalho et al. (2020), corresponds to $Z \lesssim 0.2 \, Z_\odot$. In Fig. 1, right panel, a log(C iii]$\lambda$1909/He ii$\lambda$1640) vs. log(C iv$\lambda$1549/C iii]$\lambda$1909) diagram containing our observational data (blue points) is shown. The horizontal black line defines the empirical separation between AGNs and SFs where, according to Dors et al. (2018), objects with

$$\log(\text{C III}]\lambda 1909/\text{He II}\lambda 1640) \lesssim 0.5 \quad (2)$$

are classified as AGNs, otherwise, SFs (see also Hirschmann et al. 2019). It is evident that all objects in our sample lie within the AGN zone, thereby corroborating their prior classifications.

The literature-compiled optical line intensity ratios were already corrected by Galactic and internal extinction. Different approaches to the dust effects could result in distinct reddening correction functions yielding different line intensity ratios. However, Dors et al. (2022) showed that the uncertainty in emission line ratios resulting from different approaches to the extinction correction is comparable to that caused by errors in emission line measurements ($\sim 0.1$ dex, e.g. Kennicutt et al. 2003). Regarding the UV lines, following Nagao et al. (2006), no internal reddening correction (or only Galactic extinction correction) was applied to the fluxes of the involved emission lines. This procedure was adopted because the emission lines have close wavelengths, the objects are mostly face-on and only one hydrogen line (Ly$\alpha$) is measured in the UV spectrum. It is worth mentioning that, for our sample, it is not possible to carry out reddening correction by using the He ii$\lambda$1640/$\lambda$4686 line ratio (e.g. Bergeron et al. 1981), hence for most of the objects the absolute UV emission line fluxes are not available. In any case, we noted in the studies from in which the data were compiled that the effect of extinction correction in the line ratios considered here is lower than the uncertainty ($\sim 30\%$) on them (see Ferland & Osterbrock 1987). In fact, for instance, considering the UV data for NGC 7674 (one object of our sample) from Kraemer et al. (1994), the observational C iii]$\lambda$1909/Ly$\alpha$ line ratio is $0.17 \pm 0.09$, while the reddening corrected value is $0.12 \pm 0.06$.

One concern in the present study is that the optical and UV spectra for the same object were obtained using different instruments, which may lead to distinct extraction apertures (e.g. Humphrey et al. 2008). Differences in observational setups may be sources of discrepancies in the derived parameters through optical and UV line fluxes (e.g. Berg et al. 2024). In particular, emission from nuclear H ii regions can contribute by distinct amounts to the observed UV and optical AGN spectra. For instance, Thomas et al. (2018b) fitted photoionization model results to observational emission-line intensity ratios of 2766 Seyfert galaxies, whose data were taken from the Sloan Digital Sky Survey (SDSS) data release 7 (Abazajian et al. 2009). These authors found that, even for AGNs with high ionization degree, i.e. log([O iii]$\lambda$5007/H$\beta$) $\gtrsim 0.9$, about 30 per cent of the Balmer flux arises from H ii regions (see also de Mellos et al. 2024). However, Humphrey et al. (2008), who analyzed





**Table 1.** Object names, references from where the optical and ultraviolet emission lines were compiled, redshifts and AGN classes. The logarithm of the X-ray luminosity (measured in the 14-195 keV band) values (when available) were taken from https://swift.gsfc.nasa.gov/results/bs70mon/.

| Object | Reference | | redshift | Class | log($L_x$) [erg s$^{-1}$] |
|---|---|---|---|---|---|
| | Optical | Ultraviolet | | | |
| NGC 5506 | Dopita et al. (2015) | Bergeron et al. (1981) | 0.006084[1] | Seyfert 1.9[2] | 43.31 |
| NGC 1068 | Koski (1978) | Snijders et al. (1986) | 0.003793[1] | Seyfert 2[2] | 42.05 |
| NGC 7674 | Kraemer et al. (1994) | Kraemer et al. (1994) | 0.029030[1] | Seyfert 2[3] | — |
| NGC 4507 | Phillips et al. (1983) | Bergeron et al. (1981) | 0.011801[1] | Seyfert 2[2] | 43.77 |
| Mrk 3 | Koski (1978) | Malkan & Oke (1983) | 0.013509[1] | Seyfert 2[2] | 43.76 |
| Mrk 573 | Koski (1978) | MacAlpine (1988) | 0.017212[1] | Seyfert 2[3] | — |
| Mrk 1388 | Osterbrock (1985) | MacAlpine (1988) | 0.020954[1] | Seyfert 1.9[4] | — |
| MCG-3-34-64 | Koss et al. (2021) | De Robertis et al. (1988) | 0.016541[1] | Seyfert 1[4] | 43.27 |
| III Zw77 | Ferland & Osterbrock (1987) | Ferland & Osterbrock (1987) | 0.033829[1] | Seyfert 1.5[4] | — |
| 4C+40.36 | Humphrey et al. (2008) | Humphrey et al. (2008) | 2.349[5] | radio galaxy[5] | — |
| 4C+23.56 | Humphrey et al. (2008) | Humphrey et al. (2008) | 2.479[5] | radio galaxy[5] | — |

References: (1) https://ned.ipac.caltech.edu/. (2) Koss et al. (2021), (3) Osterbrock & Martel (1993), (4) Véron-Cetty & Véron (2006), (5) Humphrey et al. (2008).

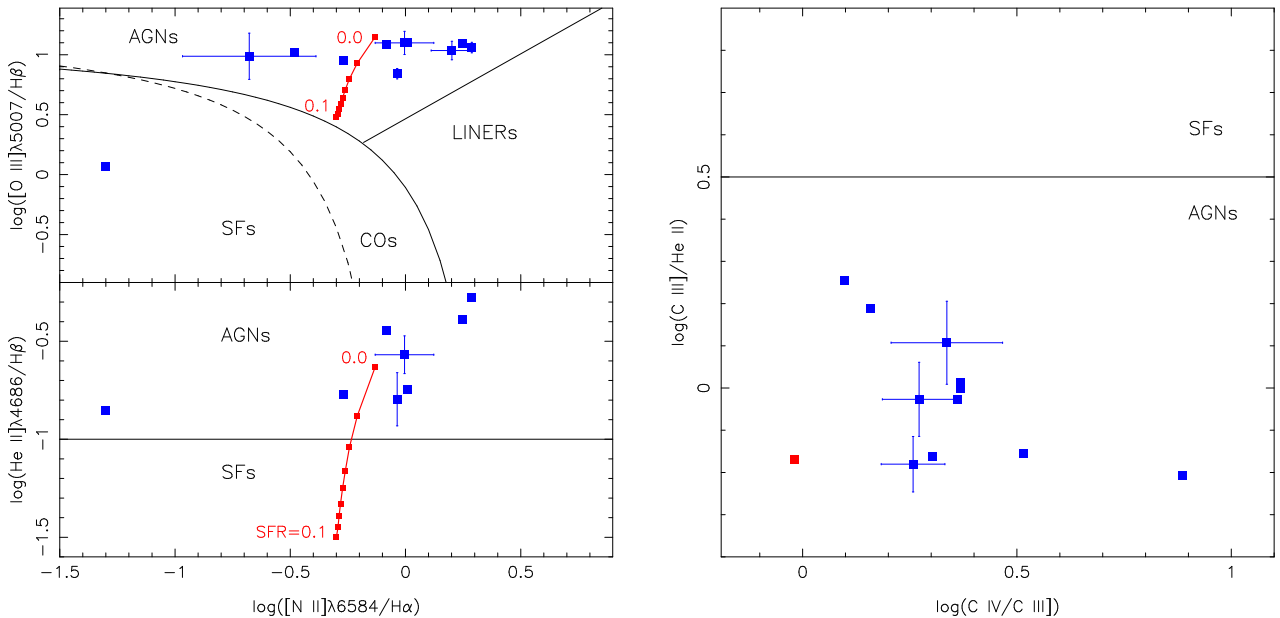

**Figure 1.** Diagnostic diagrams used to identify AGN-like and SF-like objects. Blue points represent observational data of our sample of AGN (see Sect. 2.1). Bottom left panel: He II$\lambda$4686/H$\beta$ vs. [N II]$\lambda$6584/H$\alpha$ diagram suggested by Shirazi & Brinchmann (2012). The line represents the separation criterion suggested by Molina et al. (2021) and given by Eq. 1. Upper left panel: [O III]$\lambda$5007/H$\beta$ vs. [N II]$\lambda$6584/H$\alpha$ (Baldwin et al. 1981). Dashed and solid curves represent the empirical and theoretical separation criteria suggested by Kauffmann et al. (2003) and Kewley et al. (2006), respectively. The region between the Kauffmann et al. (2003) and Kewley et al. (2006) curves is occupied by Composite Objects (COs). The solid line represents the criterion suggested by Kewley et al. (2006) to separate AGNs from low-ionization nuclear emission-line regions (LINERs). In the left panels, the red line connects the results of photoionization models simulating composite (AGN+SF) objects (see Sect. 2.1). The star formation rate (SFR, in units of M$_\odot$ yr$^{-1}$) of the simulated COs is indicated. Right panel: C III]$\lambda$1909/He II$\lambda$1640 vs. C IV]$\lambda$1549/C III]$\lambda$1909 suggested by Hirschmann et al. (2019). The line corresponds to the empirical separation criterion suggested by Dors et al. (2018) and given by Eq. 2. Composite model results are represented by only one red point, since the AGN luminosity of the UV lines considered is higher (by ∼ 0.3 dex) in comparison to those of CNSFRs (see Table 2).

the aperture effect on emission line measurements for a sample of high-$z$ galaxies, found that the size of the aperture does not influence line intensity ratios by more than 10 per cent (see also Daltabuit & Cox 1972; Kewley et al. 2005; Mannucci et al. 2021; Arellano-Córdova et al. 2022; Pilyugin et al. 2022). It is worth mentioning that the objects analyzed by Humphrey et al. (2008) are at $z \sim 2.4$, where the impact of aperture on emission line fluxes is lower compared to that on local AGNs (e.g. Davies et al. 2016; D'Agostino et al. 2019a; Pilyugin et al. 2020; Hviding et al. 2023; Molina et al. 2023).

To investigate the flux contribution from H II regions to the observed spectrum of an AGN, we built photoionization models using version 23.01 of the CLOUDY code (Ferland et al. 2017) and following a similar methodology to the one adopted by Dors et al. (2018). In particular, we considered an AGN surrounded by eight circumnuclear star-forming regions (CNSFRs) as illustrated in Fig. 2. The nebular parameters adopted in the models are summarized in what follows:

• CNSFR models: We assumed solar metallicity, electron density $N_e = 200$ cm$^{-3}$ (see, e.g. Copetti et al. 2000), spherical geometry,





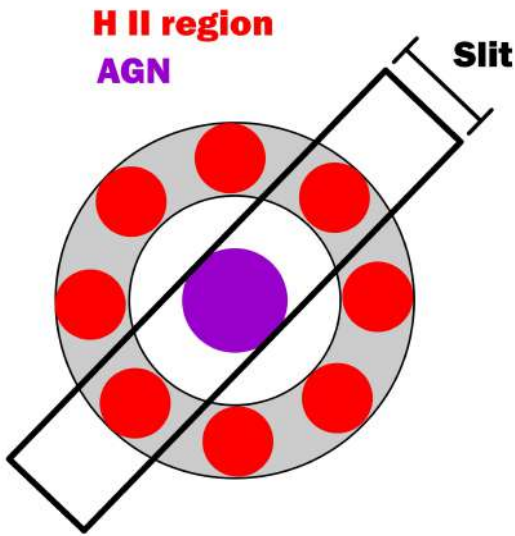

**Figure 2.** Illustration of a galaxy nuclei containing an AGN and circumnuclear star-forming regions (CNSFRs), as indicated. This scenario is used to investigate the flux contribution of H II regions to the spectrum of an AGN obtained by distinct slit apertures, as indicated (see Sect. 2.1).

inner radius of 3 pc, and the logarithm of the number of ionizing photons $\log Q(\text{H})\,[\,\text{s}^{-1}] = 51$, being about the mean value derived by Díaz et al. (2007) for CNSFRs in the NGC 2903, NGC 3351 and NGC 3504 galaxies. We considered as the ionizing source a stellar cluster instantaneously formed, with an age of 3.5 Myr, stellar mass of $10^6$ M$_\odot$, Initial Mass Function (IMF) as defined by Kroupa (2002), and a spectrum taken from the STARBURST99 code (Leitherer et al. 1999). These nebular and stellar parameters are typical of CNSFRs in nearby galaxies as derived by Díaz et al. (2007) and Dors et al. (2008).

- AGN models: We assumed an electron density $N_e = 500\,\text{cm}^{-3}$, $Z = Z_\odot$, $\log Q(\text{H})\,[\,\text{s}^{-1}] = 51$, spherical geometry, inner radius of 3 pc, and a power law with the optical to X-rays slope $\alpha_{ox} = -1.1$ (Tananbaum et al. 1979) as the ionizing source. These parameters are typical of local AGNs as derived through photoionization models simulating Narrow Line Regions (NLRs; e.g. Dors et al. 2017; Pérez-Montero et al. 2019).

In Table 2, the resulting luminosity values for the optical and UV lines involved in the diagrams presented in Fig. 1, and predicted by the pure AGN and CNSFR models, are listed.

By using the predicted pure-AGN and CNSFR luminosity values listed in Table 2, we simulated distinct slit apertures encompassing one by one each CNSFR. The resulting line intensity ratio was considered as:

$$\frac{I(\lambda_1)}{I(\lambda_2)} = \frac{L(\lambda_1)_{\text{AGN}} + n \times L(\lambda_1)_{\text{H\,II}}}{L(\lambda_2)_{\text{AGN}} + n \times L(\lambda_2)_{\text{H\,II}}}, \quad (3)$$

being $I(\lambda_1)/I(\lambda_2)$ the intensity ratio of two emission lines, $L$ the luminosity of a given emission line predicted by the photoionization models, and $n$ the number of H II regions. The value $n = 0$ represents the emission-line ratio of a pure AGN and $n = 8$ the number of CNSFRs as observed, for instance, in NGC 3351 (Díaz et al. 2007; Hägele et al. 2007, 2010). In Fig. 1, left panels, the results of our simulations in the [O III]/H$\beta$ versus [N II]/H$\alpha$ and He II/H$\beta$



**Table 2.** Logarithm of the luminosity [$L(\lambda)$; in erg s$^{-1}$] of some emission lines predicted by the pure-AGN and CNSFR photoionization models. Both types of models assume as input parameters the logarithm of the number of ionizing photons $\log Q(\text{H})\,[\,\text{s}^{-1}] = 51$, electron density equal to 500 cm$^{-3}$, spherical geometry, and solar metallicity (see Sect. 2.1).

| Line | log[$L(\lambda)$] | |
|---|---|---|
| | pure-AGN | CNSFR |
| H$\beta$ | 38.64 | 38.55 |
| He II$\lambda$4686 | 38.01 | 33.09 |
| [O III]$\lambda$5007 | 39.79 | 38.69 |
| H$\alpha$ | 39.09 | 39.00 |
| [N II]$\lambda$6584 | 38.96 | 38.67 |
| C IV$\lambda$1549 | 38.72 | 36.54 |
| He II$\lambda$1640 | 38.91 | 34.90 |
| C III]$\lambda$1909 | 38.74 | 36.69 |

versus [N II]/H$\alpha$ diagnostic diagrams are shown. In these figures, the photoionization model results are shown in terms of star formation rate (SFR), calculated through the predicted CNSFR model values of $L(\text{H}\alpha)$ and from the largely used relation by Kennicutt (1998):

$$\text{SFR}(\text{M}_\odot\,\text{yr}^{-1}) = 7.9 \times 10^{-42}\,L^{\text{T}}(\text{H}\alpha)\,[\text{erg s}^{-1}], \quad (4)$$

being $L^{\text{T}}(\text{H}\alpha) = n \times L(\text{H}\alpha)_{\text{H\,II}}$. It can be seen in left panels of Fig. 1 that the simulation results indicate that the flux contribution from CNSFRs to the spectra of our object sample could be negligible, hence models without star-formation would describe almost all our observational data. This result is expected, as the majority of the objects in our sample are AGNs characterized by a high degree of ionization [log([O III]$\lambda$5007/H$\beta$) ~ 1].

The above result is also supported by the simulation of composite (AGN+SF) objects by Thomas et al. (2018b), since our objects are located in the zone of their [O III]/H$\beta$ vs. [N II]/H$\alpha$ diagram corresponding to SF contributions to the emission-line ratio intensities of less than ~ 20 per cent (see also Agostino et al. 2021; Vidal-García et al. 2024). In Fig. 1, right panel, results from our simulation (red point) are shown in the UV diagnostic diagram. We found that the values of the C III]/He II and C IV/C III] emission-line ratios practically do not change due to flux contributions from CNSFRs. This result is expected, since the AGN luminosity of the UV lines considered is higher by a factor ranging from ~ 0.2 to ~ 0.4 dex in comparison to those from CNSFRs (see Table 2). Thus, it is able to argue that we are analyzing spectroscopic data of 'almost' pure-AGNs.

In Table 3, optical (in relation to H$\beta$=1.0) and the UV (in relation to He II$\lambda$1640=1.0) line intensity ratios for our AGN sample are listed. As it was mentioned before, only for the optical lines the reddening correction was carried out.

### 2.2 Photoinization models

For the purpose of our analysis, we built an extensive grid of photoionization models using version 23.01 of the CLOUDY code (Ferland et al. 2017) to simulate NLR of AGNs. We apply the same methodology as in Dors et al. (2022). The adopted nebular parameters are described below.

- Spectral Energy Distribution (SED): We consider SEDs parametrized by the continuum between 2 keV and 2500Å (Tananbaum et al. 1979) and described by a power law with a spectral index $\alpha_{ox}$ ranging from $-1.5$ to $-0.8$, with a step of 0.1. Results of photoionization models assuming this $\alpha_{ox}$ range reproduce optical



**Table 3.** Intensities of optical (relative to H$\beta$=1.0) and UV (relative to He II$\lambda$1640=1.0) emission lines taken from the literature (see Table 1). Only the optical line ratios are reddening corrected. The [C II]$\lambda$2325 emission line corresponds to the sum of $\lambda$2323.50, $\lambda$2324.69, $\lambda$2325.40, $\lambda$2326.93, and $\lambda$2328.12Å lines.

| Line | NGC 5506 | NGC 1068 | NGC 7674 | NGC 4507 | Mrk 3 | Mrk 573 | Mrk 1388 | MCG-3-34-64 | III Zw77 | 4C+40.36 | 4C+23.56 |
|---|---|---|---|---|---|---|---|---|---|---|---|
| | | | | | Optical (relative to H$\beta$=1.0) | | | | | | |
| [O II]$\lambda$3727 | 2.64 ± 0.03 | 1.23 | 1.29 ± 0.27 | 2.75 | 3.52 | 2.92 | 0.79 | 2.73 ± 0.03 | 0.15 | 4.26 ± 0.66 | 2.57 ± 1.24 |
| [O III]$\lambda$4363 | 0.11 ± 0.004 | 0.22 | 0.12 ± 0.03 | 0.35 | 0.24 | 0.18 | 0.69 | 0.27 ± 0.003 | 0.16 | — | — |
| H$\beta$ | 1.00 ± 0.01 | 1.00 | 1.00 | 1.00 | 1.00 | 1.00 | 1.00 | 1.00 | 1.00 | 1.00 | 1.00 |
| He II$\lambda$4686 | 0.16 ± 0.05 | 0.41 | 0.27 ± 0.06 | 0.17 | 0.18 | 0.36 | — | 0.53 ± 0.006 | 0.14 | — | — |
| [O III]$\lambda$5007 | 6.95 ± 0.04 | 12.42 | 12.55 ± 2.51 | 8.91 | 12.67 | 12.12 | 10.53 | 11.48 ± 0.12 | 1.17 | 10.85 ± 1.60 | 9.71 ± 4.20 |
| [N II]$\lambda$5755 | 0.02 ± 0.05 | 0.02 | — | — | 0.05 | — | — | — | — | — | — |
| He I$\lambda$5876 | 0.11 ± 0.18 | 0.15 | — | — | 0.08 | 0.10 | 0.09 | 0.18 ± 0.005 | 0.07 | — | — |
| H$\alpha$ | 2.86 ± 0.02 | 2.57 | 3.70 ± 0.78 | 2.83 | 3.10 | 2.95 | 2.84 | 2.86 ± 0.03 | 2.84 | 2.94 ± 0.43 | 2.85 ± 1.24 |
| [N II]$\lambda$6583 | 2.65 ± 0.02 | 4.55 | 3.68 ± 0.77 | 1.53 | 3.18 | 2.47 | 0.95 | 5.55 ± 0.05 | 0.14 | 4.70 ± 0.69 | 0.60 ± 0.32 |
| [S II]$\lambda$6716 | 0.94 ± 0.01 | 0.26 | 0.54 ± 0.11 | 0.58 | 0.73 | 0.75 | 0.21 | 0.78 ± 0.01 | 0.02 | — | — |
| [S II]$\lambda$6731 | 1.06 ± 0.01 | 0.55 | 0.64 ± 0.13 | 0.64 | 0.82 | 0.80 | 0.18 | 0.93 ± 0.01 | 0.02 | — | — |
| | | | | | Ultraviolet (relative to He II$\lambda$1640=1.0) | | | | | | |
| Ly$\alpha$ | — | 3.81 ± 0.79 | 9.15 | 13.65 | 7.11 | 12.00 | 12.94 | 5.56 | 11.11 | 14.87 ± 1.25 | 5.06 ± 0.34 |
| He II$\lambda$1640 | 1.00 | 1.00 | 1.00 | 1.00 | 1.00 | 1.00 | 1.00 | 1.00 | 1.00 | 1.00 | 1.00 |
| O III]$\lambda$1664 | — | 0.11 ± 0.04 | — | — | — | — | — | — | 0.87 | 0.16 ± 0.02 | 0.26 ± 0.03 |
| C IV$\lambda$1549 | 2.25 | 2.78 ± 0.60 | 2.23 | 2.41 | 2.33 | 2.30 | 2.17 | 1.39 | 4.77 | 1.76 ± 0.15 | 1.20 ± 0.15 |
| C III]$\lambda$1909 | 1.80 | 1.28 ± 0.27 | 1.54 | 1.03 | 1.00 | 0.70 | 0.94 | 0.69 | 0.62 | 0.94 ± 0.17 | 0.66 ± 0.08 |
| [C II]$\lambda$2325 | — | 0.28 ± 0.06 | — | 0.46 | 1.00 | — | — | — | — | 0.56 ± 0.06 | 0.52 ± 0.06 |

(e.g. Carvalho et al. 2020; Ji et al. 2020) and UV (e.g. Dors et al. 2014, 2019) narrow emission lines of AGNs.

• **Metallicity** ($Z$): We consider the metallicity in relation to the solar one ($Z/Z_\odot$) in the range of 0.2-3.0, with a step of 0.2. This metallicity range is similar to the one derived through the $T_e$-method for local Seyfert nuclei (e.g. Dors et al. 2020b).

• **Abundance relations**: We adopt the (N/O)-(O/H) abundance relation given by log(N/O) = [1.29 × (12 + log(O/H))] − 11.84 derived by Carvalho et al. (2020) and representative for local AGNs. For the C-O relation, we assumed

$$(C/H) = 6.0 \times 10^{-5} \times (Z/Z_\odot) + 2.0 \times 10^{-4} \times (Z/Z_\odot)^2 \quad (5)$$

derived for H II regions by Dopita et al. (2006). For the helium, we assumed the 12 + log(He/H) = 0.1215 × [12 + log(O/H)]$^2$ − 1.8183 × [12 + log(O/H)] + 17.6732 derived by Dors et al. (2022) by using the $T_e$-method. Other elements were linearly scaled with the metallicity. Following Nagao et al. (2006), who found that photoionization models with dust grains predict line flux ratios which are in disagreement with most of the observed values, our models are dust free. However, other studies (e.g. Groves et al. 2004; Feltre et al. 2016) include dust in their models to reproduce observations.

• **Electron density** ($N_e$): We assume $N_e$ values in the range of 100-5000 cm$^{-3}$ (step of 100 cm$^{-3}$), typical of NLRs (e.g. Zhang 2024).

In total, the grid contains 69 600 photoionization models.

### 2.3 Strong-line method

#### 2.3.1 UV calibration

We consider a semi-empirical calibration between fluxes of UV emission lines and the metallicity proposed by Dors et al. (2019). This calibration was obtained by comparing the results of photoionization models, simulating NLRs, with narrow UV emission line intensity ratios of 77 AGN with redshifts between 0 and 3.8, divided in Seyfert 2s (9 objects), high-z radio galaxies (61 objects), radio-quiet type-2 (1 object), and type-2 quasars (6 objects). The semi-empirical calibration is:

$$(Z/Z_\odot) = (2.13\pm0.09) + (2.41\pm0.19)\,x^2 + (4.76\pm0.58)\,C43^2 \\ + (7.79\pm0.59)\,x\,C43 - (4.64\pm0.19)\,x \quad (6) \\ - (5.64\pm0.48)\,C43,$$

where $C43 = \log[(C\,IV\lambda1549 + C\,III]\lambda1909)/He\,II\lambda1640]$ and $x = C3C4 = \log(C\,III]\lambda1909/C\,IV\lambda1549)$. This calibration is valid for the range of $-2 \lesssim C43 \lesssim 1$, which translates to $0.2 \lesssim (Z/Z_\odot) \lesssim 4$.

Also, Dors et al. (2019) proposed the following semi-empirical calibration to derive the ionization parameter:

$$\log U = -(0.14\pm0.02) \times C3C4^2 - (1.10\pm0.01) \times C3C4 - (1.80\pm0.01), \quad (7)$$

valid for the $-1 \lesssim C3C4 \lesssim 0.5$ range, which correspond to $-2.5 \lesssim \log U \lesssim -1.0$.

The photoionization models assumed by Dors et al. (2019) consider a fixed value of $N_e = 500$ cm$^{-3}$, a nitrogen-metallicity relation

$$(N/N_\odot) = 4.5 \times (Z/Z_\odot)^{1.2} \quad (8)$$

and a fixed value of

$$\log(C/O) = -0.50. \quad (9)$$

It is worth mentioning that Dors et al. (2019) tested various abundance relations and only the photoionization models that assumed the relations mentioned above successfully reproduced the UV emission line ratios of their sample of 77 AGNs. Although these authors had also proposed a calibration between N V$\lambda$1240/He II$\lambda$1640 and $Z$, the intensity of this nitrogen line is available for a few objects of our sample and was not considered here.

#### 2.3.2 Optical calibrations

We consider distinct calibrations between the intensities of optical emission line ratios and metallicity. The first one is the empirical calibration proposed by Dors (2021):

$$Z = (-1.00\pm0.09)P + (0.036\pm0.003)R_{23} + (8.80\pm0.06), \quad (10)$$



where $Z \equiv 12 + \log(\text{O/H})$, $R_{23}=([\text{O\,\textsc{ii}}]\lambda3727 + [\text{O\,\textsc{iii}}]\lambda4959 + \lambda5007)/\text{H}\beta$ and $\text{P}=([\text{O\,\textsc{iii}}]\lambda4959+\lambda5007/\text{H}\beta)/R_{23}$. This calibration, valid for the range of $5 \lesssim R_{23} \lesssim 20$ or $0.3 \lesssim (Z/Z_\odot) \lesssim 2.0$, was obtained by using narrow emission-line intensity ratios of a large sample of 109 AGNs (Seyfert 1s and 2s) and direct estimates of O/H based on the $T_\text{e}$-method. This calibration yields O/H values somewhat lower ($\sim 0.2$ dex) than those derived from photoionization models (a strong-line method) and, in principle, its use results in O/H values similar to those derived using the $T_\text{e}$-method.

It must be noted that P can be also written as a function of $O32=[\text{O\,\textsc{iii}}]\lambda5007/[\text{O\,\textsc{ii}}]\lambda3727$, i.e. $\text{P}=O32/(1+O32)$, which is related with the ionization parameter $U$, see discussion below. With this in mind, the Eq. 10 implies that the metallicity derived via this calibration depends on the ionization parameter as well as the $R_{23}$ index (e.g. McGaugh 1991).

The second relation is the semi-empirical calibration between $N2 = \log([\text{N\,\textsc{ii}}]\lambda6584/\text{H}\alpha)$ and $Z$ proposed by Carvalho et al. (2020). These authors compared the results of photoionization models simulating NLRs with spectroscopic data of confirmed Seyfert 2 nuclei finding

$$(Z/Z_\odot) = 4.01^{N2} - 0.07, \quad (11)$$

valid for the range of $-0.7 \lesssim N2 \lesssim 0.6$, which translates to $0.3 \lesssim (Z/Z_\odot) \lesssim 2.0$.

Carvalho et al. (2020) also proposed a semi-empirical calibration to estimate the ionization parameter given by

$$\log U = [0.57 \times (\log O32)^2] + [1.38 \times \log O32] - 3.14, \quad (12)$$

where $O32=[\text{O\,\textsc{iii}}]\lambda5007/[\text{O\,\textsc{ii}}]\lambda3727$. This calibration is valid for the range of $-1.5 \lesssim O32 \lesssim 1.0$, which translates to $-4.0 \lesssim \log U \lesssim -2.0$.

### 2.3.3 Zhu et al. calibrations

Zhu et al. (2024), by using the MAPPINGS code (Sutherland et al. 2018), built a grid of photoionization models simulating NLRs of AGNs. Briefly, these models assume plane-parallel geometry, distinct values for the gas pressure, dust content, metallicity in the range of $7.2 \lesssim 12 + \log(\text{O/H}) \lesssim 9.5$ [$0.03 \lesssim (Z/Z_\odot) \lesssim 6.5$], and $\log U$ ranging from $-3.8$ to $-1.0$. The (C/O)-(O/H) relation assumed in the models by Zhu et al. (2024) is that derived from stellar data by Nicholls et al. (2017) and represented by

$$\log(\text{C/O}) = \log(10^{-0.8} + 10^{[\log \text{O/H}+2.72]}). \quad (13)$$

The discrepancy between the emission line ratios calculated by the CLOUDY and MAPPINGS codes can differ by $\sim 0.1$ dex (D'Agostino et al. 2019b; Zhu et al. 2023).

Concerning the $C43$ calibration by Zhu et al. (2024), these authors showed that $Z$ values estimated through their calibrations are consistent with those above by a factor lower than $\sim 0.1$ dex. From the several theoretical calibrations proposed by Zhu et al. (2024), we consider the ones between the $C3C4$ and $O32$ lines ratios and $\log U$, whose equations are presented in Table 5 of Zhu et al. (2024).

The uncertainty in O/H abundance estimates obtained via strong-line methods, calculated from the errors of observational emission-line fluxes and the fitting coefficients of calibrations, is in order of $\sim 0.2$ dex, which translates into an uncertainty of $\sim 50$ per cent in $Z$ estimates (e.g. Kobulnicky et al. 1999; Denicoló et al. 2002; Dors et al. 2017; Zhu et al. 2024). The uncertainty in $U$ estimates derived from the comparison between photoionization mod-



els and AGN observational data is $\sim 0.15$ dex (e.g. Pérez-Montero et al. 2019; Dors et al. 2019).

The uncertainty values above are considered in the $Z$ and $U$ estimates from the strong emission line methods assumed in the present study.

### 2.4 $T_\text{e}$-method

We estimated, for each object of our sample of AGNs, the total abundance of the oxygen and carbon (in relation to hydrogen) through the $T_\text{e}$-method and by using the 1.1.13 version of the PYNEB code (Luridiana et al. 2015).

#### 2.4.1 Oxygen abundance

The ionic abundances for the oxygen were calculated according to:

$$\frac{\text{O}^+}{\text{H}^+} = f\left(\frac{[\text{O\,\textsc{ii}}]\lambda3727}{\text{H}\beta}, T_\text{low}, N_\text{e}\right) \quad (14)$$

and

$$\frac{\text{O}^{2+}}{\text{H}^+} = f\left(\frac{[\text{O\,\textsc{iii}}]\lambda5007}{\text{H}\beta}, T_\text{high}, N_\text{e}\right). \quad (15)$$

Values for $N_\text{e}$ and $T_\text{high}$ were obtained from the [S\,\textsc{ii}]$\lambda6716/\lambda6731$ and [O\,\textsc{iii}]$(\lambda4959 + \lambda5007)/\lambda4363$ line ratios, respectively, where the theoretical ratio [O\,\textsc{iii}]$(\lambda4959/\lambda5007)$=0.33 (Storey & Zeippen 2000) was considered. The value for $T_\text{low}$ was obtained either from [N\,\textsc{ii}]$(\lambda6548 + \lambda6584)/\lambda5755$ (3/11 objects) or from the theoretical relation proposed by Dors et al. (2020b):

$$t_\text{low} = (\text{a} \times t_\text{high}^3) + (\text{b} \times t_\text{high}^2) + (\text{c} \times t_\text{high}) + \text{d}, \quad (16)$$

where a = 0.17, b = $-1.07$, c = 2.07 and d = $-0.33$, while $t_\text{low}$ and $t_\text{high}$ represent $T_\text{low}$ and $T_\text{high}$, respectively, in units of $10^4$ K.

To calculate the total oxygen abundance (O/H) we adopted the following approach:

$$\frac{\text{O}}{\text{H}} = \text{ICF}(\text{O}^+ + \text{O}^{2+}) \times \left[\frac{\text{O}^+}{\text{H}^+} + \frac{\text{O}^{2+}}{\text{H}^+}\right]. \quad (17)$$

To derive the oxygen ionization correction factor, we assume the usual approach (see Izotov et al. 2006; Flury & Moran 2020; Dors et al. 2022):

$$\text{ICF}(\text{O}^+ + \text{O}^{2+}) = \frac{\text{He}^+ + \text{He}^{2+}}{\text{He}^+}, \quad (18)$$

being $\text{He}^+/\text{H}^+$ and $\text{He}^{2+}/\text{H}^+$ the ionic abundance ratios derived from the observational He\,\textsc{i}$\lambda5876/\text{H}\beta$ and He\,\textsc{ii}$\lambda4686/\text{H}\beta$ line intensity ratios, respectively, measured in the spectrum of each object (see Table 3). For objects for which it is not possible to derive the $\text{ICF}(\text{O}^+ + \text{O}^{2+})$ due to the lack of helium line measurements (see Table 3), a mean ICF value equal to 1.5, as derived by Dors et al. (2022) for a large sample of local AGN, was considered.

#### 2.4.2 Carbon abundance

The carbon abundance estimates, based on UV emission lines, suffer large uncertainties due to the need of using the Ly$\alpha$ as a reference line, hence it is a resonance line undergoing a complex radiative transfer (e.g. Laursen et al. 2009) and it could be blended with geocoronal hydrogen emission (e.g. Bergeron et al. 1981). Due to these





**Table 4.** Physical parameters for our sample. $T_{O^{2+}}$ is the electron temperature calculated directly from the observational value of the [O III]($\lambda 4959+\lambda 5007$)/$\lambda 4363$ by using the PYNEB code. $T_{C^{3+}}$ and $T_{C^{2+}}$ are temperatures calculated from the theoretical relations given by the Eqs. 24 and 25, respectively, while $T_{O^+}$ from Eq. 16. Ionic abundances are derived as described in Sect. 2.4. The electron density is derived from the [S II]$\lambda 6716$/$\lambda 6731$ by using the PYNEB code. ICFs and total abundance values are estimated as described in Sect. 2.4. Metallicity ($Z/Z_{\odot}$) values are estimated through distinct methods, according to the following sub-indexes: $D20$ via $T_{\rm e}$-method (Dors et al. 2020b); $D19$ via the calibration by Dors et al. (2019); $D21$ via the calibration by Dors (2021); $C20$ via the semi-empirical calibration by Carvalho et al. (2020). $O32$ and $C3C4$ corresponds to [O III]$\lambda 5007$/[O II]$\lambda 3727$ and [C III]$\lambda 1909$/[C IV]$\lambda 1549$ line ratios, respectively, used to estimate the logarithm of the ionization parameters through the calibrations proposed by $C20$, $Z24$ and $D19$. The error in $Z$ estimates is attributed to be 50 per cent and in $\log U$ estimates 0.15 dex (see Sect. 2.3).

| | NGC 5506 | NGC 1068 | NGC 7674 | NGC 4507 | Mrk 3 | Mrk 573 | Mrk 1388 | MCG-3-34-64 | III Zw77 | 4C+40.36 | 4C+23.56 |
|---|---|---|---|---|---|---|---|---|---|---|---|
| $T_{C^{3+}}$ | 14567 | 14719 | 12004 | 23174 | 15743 | 14188 | — | 17400 | — | — | — |
| $T_{C^{2+}}$ | 12694 | 12803 | 10757 | 17633 | 13516 | 12420 | — | 14600 | — | — | — |
| $T_{O^{2+}}$ | 13692 | 13835 | 11310 | 22157 | 14804 | 13336 | — | 16394 | — | — | — |
| $T_{O^+}$ | 7834 | 5804 | 8884 | 8527 | 9314 | 9307 | — | 9368 | — | — | — |
| $N_{\rm e}$ (cm$^{-3}$) | 1115 | 25025 | 1324 | 1030 | 1098 | 910 | 348 | 1350 | 712 | — | — |
| $\log(\text{O}^+/\text{H}^+)$ | $-3.81$ | $-3.54$ | $-4.00$ | $-3.60$ | $-3.70$ | $-3.77$ | — | $-3.78$ | — | — | — |
| $\log(\text{O}^{2+}/\text{H}^+)$ | $-4.03$ | $-3.78$ | $-3.53$ | $-4.39$ | $-3.85$ | $-3.75$ | — | $-4.01$ | — | — | — |
| $\log(\text{He}^+/\text{H}^+)$ | $-1.11$ | $-1.01$ | — | — | $-1.25$ | $-1.15$ | — | $-0.90$ | — | — | — |
| $\log(\text{He}^{2+}/\text{H}^+)$ | $-1.86$ | $-1.44$ | $-1.64$ | $-1.80$ | $-1.80$ | $-1.51$ | — | $-1.30$ | — | — | — |
| $\log(\text{C}^{2+}/\text{He}^{2+})$ | $-1.77$ | $-1.94$ | $-1.33$ | $-3.01$ | $-2.21$ | $-2.11$ | — | $-2.61$ | — | — | — |
| $\log(\text{C}^{3+}/\text{He}^{2+})$ | $-2.60$ | $-2.55$ | $-1.87$ | $-4.34$ | $-2.88$ | $-2.49$ | — | $-3.49$ | — | — | — |
| $\log(\text{C}^{2+}/\text{H}^+)$ | $-3.63$ | $-3.38$ | $-2.97$ | $-4.81$ | $-4.01$ | $-3.62$ | — | $-3.91$ | — | — | — |
| $\log(\text{C}^{3+}/\text{H}^+)$ | $-4.46$ | $-3.99$ | $-3.51$ | $-6.14$ | $-4.68$ | $-4.00$ | — | $-4.79$ | — | — | — |
| ICF(O) | 1.17 | 1.37 | 1.50* | 1.50* | 1.27 | 1.43 | — | 1.37 | — | — | — |
| ICF(C) | 2.50 | 2.27 | 1.50 | 1.30 | 2.08 | 1.85 | — | 1.75 | — | — | — |
| $12+\log(\text{O}/\text{H})$ | 8.46 | 8.79 | 8.76 | 8.63 | 8.63 | 8.69 | — | 8.55 | — | — | — |
| $12+\log(\text{C}/\text{H})_{\rm Eq.33}$ | 8.82 | 9.06 | 9.31 | 7.32 | 8.38 | 8.79 | — | 8.38 | — | — | — |
| $\log(\text{C}/\text{O})_{\rm Eq.33}$ | 0.36 | 0.27 | 0.55 | $-1.31$ | $-0.25$ | 0.10 | — | $-0.17$ | — | — | — |
| Metallicity | | | | | | | | | | | |
| $Z/Z_{\odot}(T_{\rm e})_{D20}$ | 0.59 | 1.26 | 1.17 | 0.83 | 0.83 | 1.00 | — | 0.72 | — | — | — |
| $Z/Z_{\odot}(C43)_{D19}$ | 0.47 | 0.70 | 0.54 | 0.97 | 1.01 | 1.67 | 1.11 | 1.69 | 1.50 | 1.11 | 1.77 |
| $Z/Z_{\odot}(R_{23})_{D21}$ | 0.57 | 0.65 | 0.67 | 0.66 | 1.03 | 0.88 | 0.49 | 0.81 | — | 1.02 | 0.68 |
| $Z/Z_{\odot}(N2)_{C20}$ | 0.88 | 1.34 | 0.92 | 0.62 | 0.94 | 0.82 | 0.44 | 1.42 | 0.09 | 1.25 | 0.32 |
| Ionization parameter | | | | | | | | | | | |
| $\log U\ (O32)_{C20}$ | $-2.45$ | $-1.17$ | $-1.22$ | $-2.28$ | $-2.19$ | $-2.06$ | $-0.86$ | $-2.05$ | $-1.45$ | $-2.48$ | $-2.15$ |
| $\log U\ (O32)_{Z24}$ | $-2.58$ | $-1.61$ | $-1.66$ | $-2.48$ | $-2.40$ | $-2.31$ | $-1.44$ | $-2.29$ | $-2.04$ | $-2.59$ | $-2.44$ |
| $\log U\ (C3C4)_{D19}$ | $-1.69$ | $-1.44$ | $-1.62$ | $-1.41$ | $-1.41$ | $-1.26$ | $-1.41$ | $-1.47$ | $-0.93$ | $-1.51$ | $-1.52$ |
| $\log U\ (C3C4)_{Z24}$ | $-2.07$ | $-1.77$ | $-1.98$ | $-1.69$ | $-1.69$ | $-1.44$ | $-1.68$ | $-1.66$ | $-1.00$ | $-1.77$ | $-1.70$ |

caveats, we estimated the ionic abundances by adopting a distinct methodology from the one used by previous studies (e.g. Garnett et al. 1995a; Berg et al. 2019), in which the C/O abundance is calculated through the C III$\lambda 1909$/O III$\lambda 1664$. In the present study, we estimated the C$^{2+}$ and C$^{3+}$ ionic abundances in relation to the He$^+$ derived from the C III$\lambda 1909$/He II$\lambda 1640$ and C IV$\lambda 1549$/He II$\lambda 1640$ line ratios, respectively, combined with the He$^{2+}$/H$^+$ optical abundance estimates. In this regard, initially, we estimated the carbon ionic abundances as

$$\frac{\text{C}^{2+}}{\text{He}^{2+}} = \frac{F(\text{C III}\lambda 1909)}{F(\text{He II}\lambda 1640)} \times \frac{\epsilon_{\lambda 1640}(N_{\rm e}, T_{\rm e})}{\epsilon_{\lambda 1909}(N_{\rm e}, T_{\rm e})} \quad (19)$$

and

$$\frac{\text{C}^{3+}}{\text{He}^{2+}} = \frac{F(\text{C IV}\lambda 1549)}{F(\text{He II}\lambda 1640)} \times \frac{\epsilon_{\lambda 1640}(N_{\rm e}, T_{\rm e})}{\epsilon_{\lambda 1549}(N_{\rm e}, T_{\rm e})} \quad (20)$$

where $F$ and $\epsilon$ are the measured flux and the emissivity, respectively, of the indicated emission lines. Using the PYNEB code, assuming an electron density of $N_{\rm e} = 500\,\text{cm}^{-3}$ (typical of NLRs, e.g. Dors et al. 2020a; Zhang 2024) and a range of electron temperature from 5000

K to 25 000 K, we obtained

$$\frac{\epsilon_{\lambda 1640}}{\epsilon_{\lambda 1909}} = 0.05 \times (t_{\rm high}^{-7.00}) \quad (21)$$

and

$$\frac{\epsilon_{\lambda 1640}}{\epsilon_{\lambda 1549}} = 0.03 \times (t_{\rm vhigh}^{-8.79}). \quad (22)$$

where $t_{\rm high}$ and $t_{\rm vhigh}$ are the electron temperature of the high and very high ionization zones, respectively, in units of $10^4$ K. Then, the total carbon abundance in relation to the hydrogen is obtained by

$$\frac{\text{C}}{\text{H}} = \left(\frac{\text{He}^{2+}}{\text{H}^+}\right)_{\rm op.} \times \text{ICF}(\text{C}^{2+}+\text{C}^{3+}) \times \left[\frac{\text{C}^{2+}}{\text{He}^{2+}} + \frac{\text{C}^{3+}}{\text{He}^{2+}}\right], \quad (23)$$

where the (He$^{2+}$/H$^+$)$_{\rm op.}$ is the abundance ratio derived from optical lines and by using the $T_{\rm e}$-method.

The temperature supposition for distinct carbon ions is not as direct as for the oxygen (e.g. Hägele et al. 2008). According to the four-zone ionization model proposed by Berg et al. (2021), the gas structure in H II regions can be divided into low, intermediate, high, and very high ionization zones. Under this scenario, C$^{3+}$ (PI=64.49 eV) occupies preferably the very high ionization





zone, whose temperature is here defined by $T_{\rm vhigh}$. Direct estimates of this temperature can be obtained through a ratio between lines emitted by the $Ne^{2+}$ ion (PI=63.45 eV), i.e. [Ne III]$\lambda 3342/\lambda 3868$, which is unfortunately not available for our data sample. The $C^{2+}$ ion (PI=47.88 eV) partially spans on both intermediate and high ionization zones, thus, the temperature for this ion is similar to the one for $O^{2+}$ (PI=35.12 eV, see Garnett et al. 1995a).

As shown by Dors et al. (2020b), AGNs and H II regions tend to present distinct temperature structures, thus, not necessarily the four zones model proposed by Berg et al. (2021) is valid for our analysis. To test its validity, we use the electron temperature values for the $C^{3+}$, $C^{2+}$ and $O^{2+}$ ions predicted by the AGN photoionization model results described above. We assume the mean temperature values calculated over the nebular AGN radius times the electron density. In Fig. 3, the model predicted temperatures (in units of $10^4$ K) for $C^{3+}$ and $C^{2+}$ versus those for $O^{2+}$ are plotted in terms of $\log U$. The dashed line indicates the equality between the estimates and the error bars the typical electron temperature uncertainty of $\pm 1000$ K (e.g. Dors et al. 2023). It can be seen that the models, assuming this estimation for the temperature uncertainty, predict $T_e(C^{3+})$ values very similar to $T_e(O^{2+})$ ones, although a slight deviation in comparison to the one-by-one relation is noted. $T_e(C^{2+})$ is systematically lower than $T_e(O^{2+})$ for $O^{2+}$ temperature values higher than $\sim 20000$ K. A fitting to the model temperature predictions in Fig. 3 provide:

$$y = -0.212(\pm 0.003)x^2 + 1.342(\pm 0.008)x - 0.171(\pm 0.006) \quad (24)$$

and

$$w = -0.055(\pm 0.003)x^2 + 1.213(\pm 0.008)x - 0.101(\pm 0.006), \quad (25)$$

where $y = t_e(C^{3+})$, $x = t_e(O^{2+})$ and $w = t_e(C^{2+})$. We use the relations above to estimate the temperatures needed to infer the carbon ionic abundances.

Regarding the ICF for the carbon, Garnett et al. (1995a), by using photoionization models simulating H II regions, proposed (see also Berg et al. 2016; Skillman et al. 2020) that (C/O)=ICF($C^{2+}/O^{2+}$) × ($C^{2+}/O^{+2}$), where ICF($C^{2+}/O^{2+}$)=($O^{2+}/O$)/($C^{2+}/C$). However, we aim to estimate C/H; therefore, a distinct approach is required. We followed the approach performed by Ji et al. (2024), in which the carbon ICF was obtained by comparing the results of photoionization models with observational line intensity ratios. In Fig. 4, a log($C43$) versus log(C III/C IV) diagram containing the observational data of our sample (red triangles) are compared to the model results (circles). The color bar indicates the metallicity (in relation to solar) of the models, while the arrow indicates the direction in which the ionization parameter increases. To derive the carbon ICF for the objects for which it was possible to apply the $T_e$-method (see Table 4), we adopted the following methodology.

- First, from the comparison in Fig. 4, we selected a model that better represent a given object assuming the minimal value of $\chi^2$, being

$$\chi^2 = [(C43)_{\rm mod.} - (C43)_{\rm obs.}]^2 + [(C3C4)_{\rm mod.} - (C3C4)_{\rm obs.}]^2, \quad (26)$$

where the 'mod.' and 'obs.' indexes represent the model predicted and observational line ratio, respectively.

- After the model selection, we consider the predicted ionic abundances calculated over the nebular radius times electron density. Thus, the ICF is derived by

$$\text{ICF}(C^{2+} + C^{3+}) = \frac{C}{C^{2+} + C^{3+}}. \quad (27)$$

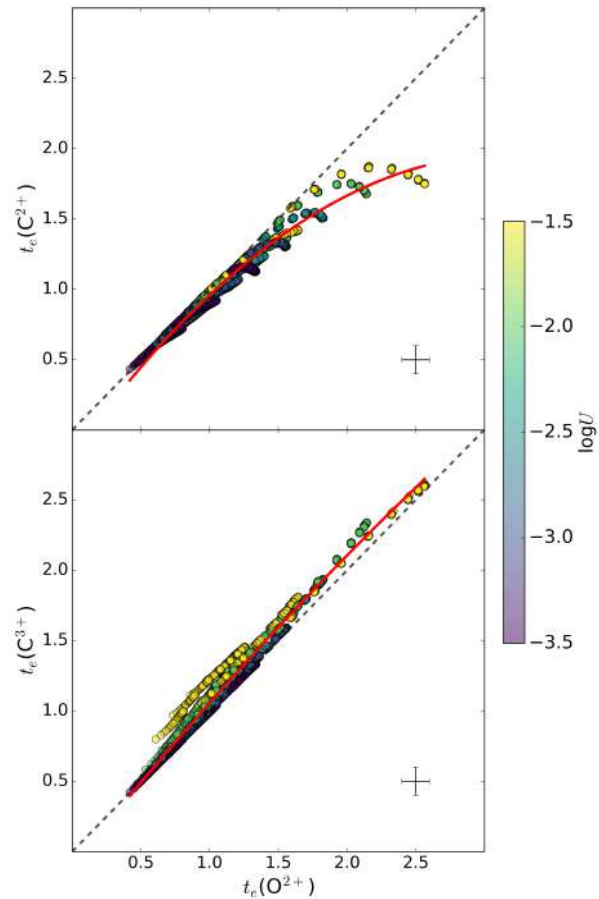

**Figure 3.** Comparisons between different electron temperature values predicted by our photoionization model results (see Sect. 2.2) simulating NLRs. The values correspond to the model-predicted mean temperature (in units of $10^4$ K) for each ion over the nebular AGN radius times the electron density. The dashed line corresponds to the equality between the temperatures. Error bars represent the typical temperature uncertainty of $\pm 1000$ K (e.g. Dors et al. 2023). The color bar represents the logarithm of the ionization parameter [$\log U$] assumed in the photoionization models. Bottom panel: $t_e(C^{3+})$ vs. $t_e(O^{2+})$. The red line represents a fitting to the points given by Eq. 24. Top panel: $t_e(C^{2+})$ vs. $t_e(O^{2+})$. The red line represents a fitting to the points given by Eq. 25.

We derive the carbon ICF that ranges from $\sim 1.3$ to $\sim 2.5$, as listed in Table 4. Furthermore, we verify the dependence of the (C/H)-$Z$ relation on the ICF estimations. With this aim, we run two new grids of photoionization models (not shown): one assuming a fixed value of $\log(C/O)_\odot = -0.30$, and another using the relation (Eq. 13) proposed by Nicholls et al. (2017). We find almost the same ICF values as those derived from our grid (see Sect. 2.2), assuming the (C/H)-$Z$ relation (Eq. 5) proposed by Dopita et al. (2006). For a detailed discussion on how uncertainties in ICFs can affect C/O estimates see, for instance, Gutkin et al. (2016).

The values adopted as references for the oxygen and carbon abundances are the solar ones: $\log(O/H)_\odot = -3.31$ and $\log(C/H)_\odot = -3.61$ derived by Allende Prieto et al. (2001) and Allende Prieto et al. (2002), respectively. As usual, we use metallicity ($Z$) and oxygen abundance [12+log (O/H)] interchangeably (see Nicholls et al. 2017 and Kewley et al. 2019 for a discussion), being its relation

$$12 + \log(O/H) = 12 + \log[(Z/Z_\odot) \times 10^{-3.31}]. \quad (28)$$





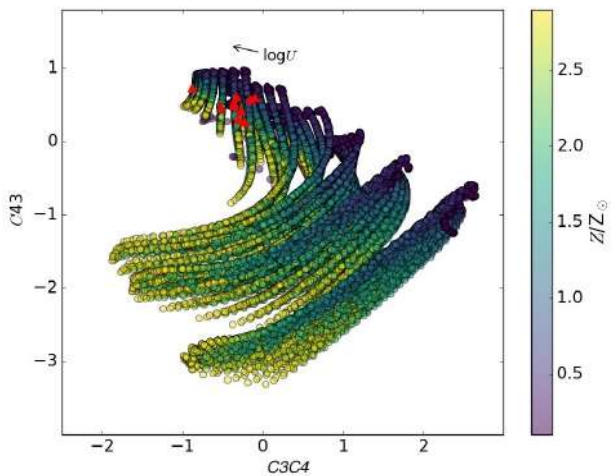

**Figure 4.** $C43=\log[(\text{C\,{\sc iv}}]\lambda1549+\text{C\,{\sc iii}}]\lambda1909)/\text{He\,{\sc ii}}\lambda1640]$ versus $C3C4=\log[\text{C\,{\sc iii}}]\lambda1909/\text{C\,{\sc iv}}]\lambda1549]$. Circles represent the results of our photoionization models (see Sect. 2.2), while red triangles the observational data of our sample (see Sect. 2.1). The color bar indicates the metallicity (in relation to the solar one) assumed in the models. The arrow indicates the direction in which the ionization parameter assumed in the models increases.

Since most of the objects in our sample do not have errors in the compiled emission lines, we adopted a typical error of 10 per cent for strong emission-lines (e.g. [O {\sc iii}]$\lambda$5007) and 20 per cent for auroral lines (e.g. [O {\sc iii}]$\lambda$4363), such as derived by Kraemer et al. (1994). This translates into an uncertainty of $\sim$ 0.1 dex in the abundance values.

For 4/11 objects of our sample is not possible to apply the $T_e$-method either because no auroral lines were measured (4C+40.36 and 4C+23.56) or due to the $RO3=[\text{O\,{\sc iii}}](\lambda 4959 + \lambda 5007)/\lambda 4363$ line ratio being out of the validity range of its relation with the electron temperature (III Zw77 and Mrk 1388) indicating temperatures higher than 25 000 K.

## 3 RESULTS & DISCUSSION

### 3.1 Metallicity discrepancy

It seems that the first metallicity estimate for NLRs through ultraviolet lines was carried out by Nagao et al. (2006), who derived sub-solar metallicity values [$0.2 \lesssim (Z/Z_\odot) \lesssim 1.0$] for selected quasars ($1.5 \leq z \leq 3.7$) by comparing narrow observational UV line intensity ratios with photoionization model results and assuming that they are mainly emitted by low-density gas clouds ($N_e \leq 10^3\,\text{cm}^{-3}$). Also, Nagao et al. (2006) did not find any evidence of the evolution of the gas metallicity in the NLRs within their studied redshift range[4] (see also Jiang et al. 2007; Matsuoka et al. 2009; De Rosa et al. 2014; Xu et al. 2018; Tang et al. 2019; Onoue et al. 2020). This result implies that the main epoch of star formation in the host galaxies of AGNs likely occurred at $z > 4$. Recent estimates of the metallicity of AGNs at $z > 6$, mainly based on UV lines, have yielded not-expected high values, around 0.3 times the solar metallicity (e.g. Schmidt et al. 2021; Isobe et al. 2023; D'Antona et al. 2023; Castellano et al. 2024; Ji et al. 2024; Rizzuti et al. 2024), supporting the result obtained by Nagao et al. (2006), in which AGNs reach their chemical maturity in the early epoch of the Universe.

---
[4] For an opposite result for local ($z < 0.5$) AGNs see Carr et al. (2023).

Another important result concerns the relation between AGN metallicity ($Z_{\text{AGN}}$) and the mass of the host stellar galaxy mass ($M_\star$). From UV emission lines, Dors et al. (2019) found a clear trend for this relation, showing that $Z_{\text{AGN}}$ increases with $M_\star$ for objects within the $0 < z < 3.8$ range. It is worth to be mentioned that this result is not observed for most local ($z < 0.4$) active nuclei with $Z_{\text{AGN}}$ estimates via optical lines (e.g. Dors et al. 2020a; Li et al. 2024; Oliveira et al. 2024). Since the optical and UV metallicity estimates in the literature have been conducted for galaxies within a similar mass range ($10^{9.5} \lesssim [M_\star/M_\odot] \lesssim 10^{12}$], although for distinct observational samples, the inconsistency in the $Z_{\text{AGN}}$-$M_\star$ relation described above does raise questions about the reliability of $Z_{\text{AGN}}$ estimates across different wavelength ranges.

The main goal of this study is to investigate the discrepancy between the metallicity of NLRs derived through UV and optical emission lines. In particular, we focus our analysis on UV metallicities based on carbon lines, specifically, from the $C43$ index, suggested by Dors et al. (2014) as a metallicity tracer. This line ratio, in principle, is a more reliable $Z$ tracer than other indexes based on lines emitted by only an ion (e.g. O {\sc iii}]$\lambda$1666/He {\sc ii}$\lambda$1640, see Zhu et al. 2024) because it involves lines emitted by ions with distinct ionization stages that occupy a large region along the nebular radius, similar to the $R_{23}$ index largely used in optical $Z$ estimates (Pagel et al. 1979).

We start our analysis by comparing metallicity estimates for our sample (see Table 1) via the $C43 - Z$ relations derived by Dors et al. (2019, Eq. 6) with those from the $T_e$-method (see Sect. 2.4). This comparison is shown in the bottom part of the panel (a) of Fig. 5, where the line represents the equality between the $Z$ values. In the top part of the panel (a) of Fig. 5, the ratio ($ZR$=y/x) between the $Z$ estimations, its mean value (<$ZR$>), a linear fitting (in red) to the points, and a line representing $ZR = 1$ are presented. It can be noted that, although the analysis is feasible for a few objects (for which the $T_e$-method could be applied) presenting a large dispersion, a clear negative trend for $ZR$ arises. The mean value derived for the metallicity ratio (<$ZR$>) is close to one and $C43$ tends to produce lower $Z$ values than those via the $T_e$-method for ($Z/Z_\odot$) $\gtrsim$ 1, while an opposite behavior is observed for ($Z/Z_\odot$) $\lesssim$ 1.

Since for the present study we were only able to compile from the literature a reduced sample of objects with the required observational data, and it was not possible to apply the $T_e$-method for all of them (only on 7/11 objects), we consider as a reference method the empirical $R_{23} - Z$ calibration (Eq. 10) proposed by Dors (2021). Comparison between $Z$ estimates via the $C43$ index and that via the $R_{23}$ is shown in panel (b) of Fig. 5. For III Zw77 was not possible to estimate $Z$ via $R_{23}$ because its line ratio intensities are out of the validity of the metallicity calibrations assumed. Fig. 5 shows that metallicity values from the $C43$ calibration by Dors et al. (2019) overpredicts $Z$ by a mean factor <$ZR$>$\sim$ 1.5. As in the case of the comparison with the results obtained through the $T_e$-method, there is a similar trend of $Z$ estimates via $C43$ and $R_{23}$.

The $Z$ estimates from the $R_{23}$ calibration and the $T_e$-method are independent of any carbon-oxygen relation, as any other calibration not involving carbon emission lines. This is evidenced by the similar behavior (see Fig. 5) of the $ZR$ ratio between estimates via $C43$ and those from $T_e$-method and $R_{23}$ as well as their similar negative trend. As an additional test of this hypothesis, in Fig. 6, $Z$ values derived from the $C43$ calibration proposed by Dors et al. (2019) are compared with those from the $N2$ calibrations proposed by Carvalho et al. (2020). Again, the same negative trend is noted. The behavior shown by these comparisons indicates that the results in Figs. 5 and 6 could be attributed mainly to the assumption of an





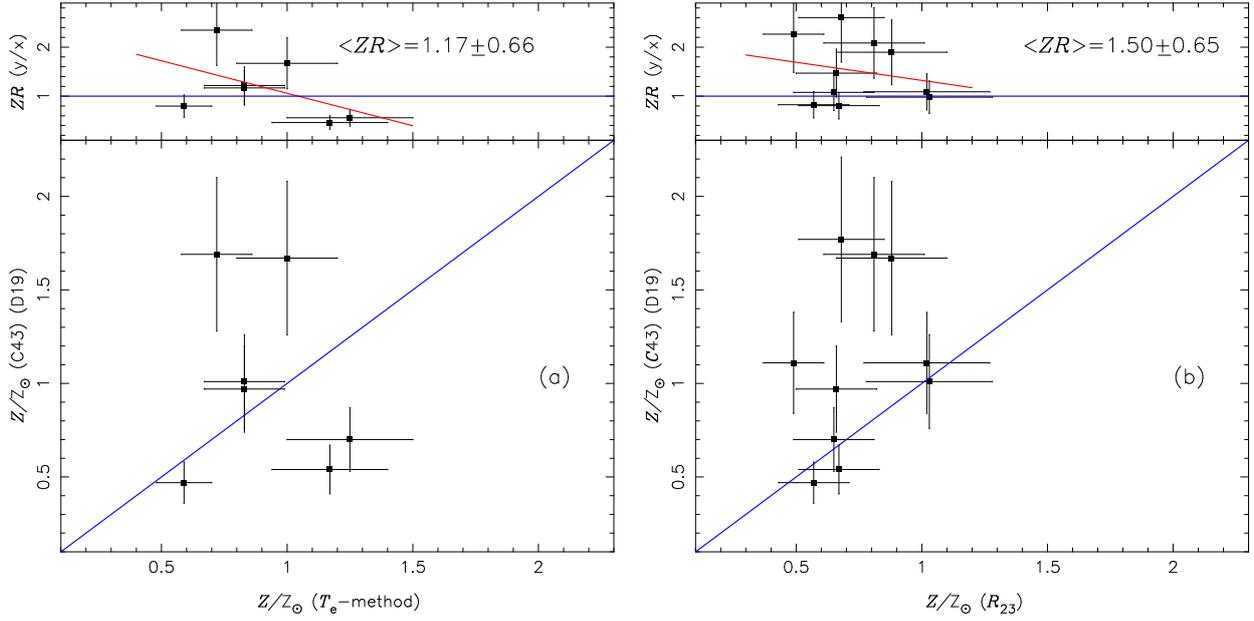

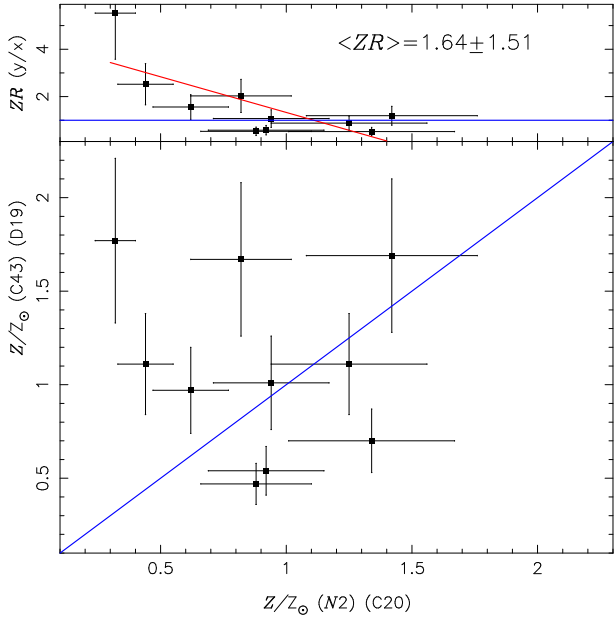

**Figure 5.** Bottom parts of panels: Comparison between metallicity $Z$ (in relation to $Z_\odot$) derived using the methods described in Sect. 2.3. Lines represent the equality between the estimates. Points represent $Z$ estimates for our sample of objects (see Table 4). $y$-axes represent estimates obtained assuming UV emission line ratios and $C43$-$Z$ calibrations, where D19 refers to estimates from the semi-empirical calibration by Dors et al. (2019, lower panels). $x$-axes represent estimates through the $T_e$-method (Dors et al. 2020b, left panels) and the empirical $R_{23}$-$Z$ calibration proposed by Dors (2021, right panels) using optical lines. Top parts of panels: Ratio ($ZR$=y/x) between the estimates. Blue lines represent the ratio $ZR$=1, while red lines represent linear fittings to the points (coefficients not shown).

**Figure 6.** As Fig. 5 but assuming $Z$ estimates via $N2$-$Z$ calibrations in the $x$-axis, being $N2$=[N II]$\lambda6584$/H$\alpha$. C20 refers to semi-empirical calibration by Carvalho et al. (2020).

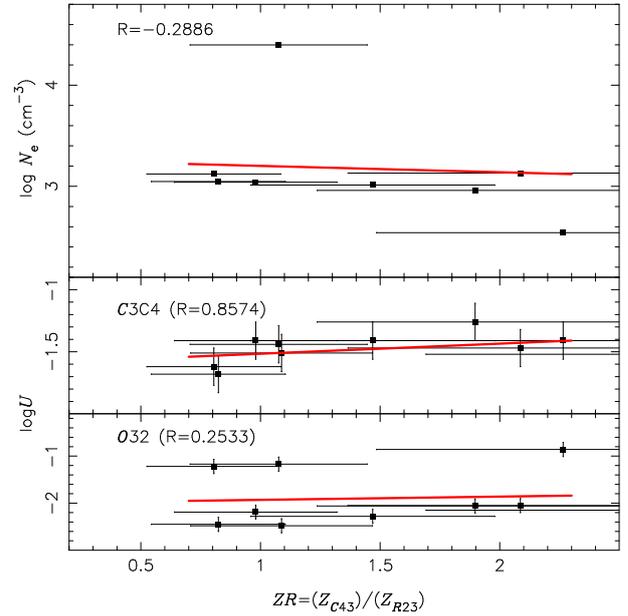

**Figure 7.** Logarithm of the ionization parameter (log $U$; bottom and middle panels) and of the electron density (log $N_e$; top panel) versus the ratio between the metallicity ($ZR$) derived by using the $C43$ and $R_{23}$ indexes. In the bottom panel, log $U$ was estimated via the $O32$ and using the Eq. 12 proposed by Carvalho et al. (2020). In the middle panel, log $U$ was estimated via the $C3C4$ and using the Eq. 7 by Dors et al. (2019). $Z_{C43}$ represents the metallicity estimated by using the calibrations by Dors et al. (2019), whilst $Z_{R23}$ by using the calibration by Dors (2021). In each panel, the red line represents a linear fitting (coefficients not shown) to the points and its corresponding value for the Pearson correlation coefficient R is indicated.

unrealistic (C/O)-(O/H) abundance relation in the models, rather than to other nebular parameters (e.g., SED or dust content).

### 3.2 Discrepancy source

To investigate the source of the discrepancy in $Z$ derived from UV and optical emission lines, we analyzed the dependence of the





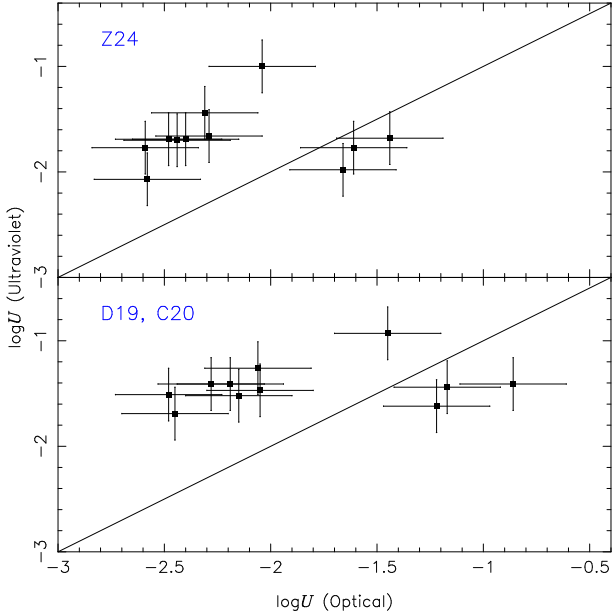

**Figure 8.** Logarithm of the ionization parameter ($\log U$) derived from $C3C4$ versus those via $O32$. Bottom panel: The calibrations proposed by Carvalho et al. (2020) and Dors et al. (2019) were used to derive $\log U$, whilst in the top panel by Zhu et al. (2024). The lines indicate the equality between the estimates.

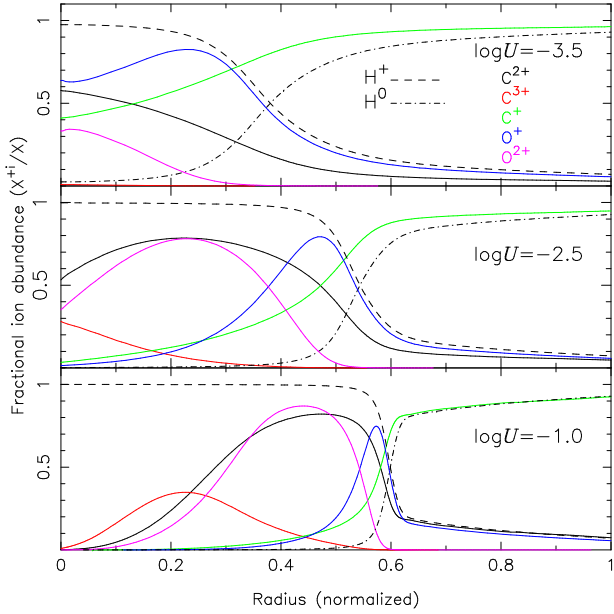

**Figure 9.** Ionic abundance fraction for distinct ions ($X^i/X$) versus the nebular radius (normalized by the outermost radius) predicted by photoionization models simulating NLRs of AGNs. The nebular parameters assumed in the models are: spherical geometry, solar metallicity, $\alpha_{ox} = -1.1$, and three values for the $\log U = -1.0, -2.5$ and $-3.5$, as indicated.

ratio ($ZR$) between $Z$ estimates obtained using the $C43$ and $R_{23}$ calibrations (see Fig. 5) on some nebular parameters. This analysis was necessary due to the limited number (7/11) of direct metallicity estimates available for our sample.

In Fig. 7, the logarithm of the ionization parameter ($\log U$) for our sample (see Table 4) derived through $O32$ (bottom panel) and $C3C4$ (middle panel) as well as the electron density ($N_e$; top panel) are plotted against the ratio ($ZR$) between $Z$ estimates via $C43$ and $R_{23}$. Values of $\log U$ estimated from $O32$ and $C3C4$ are derived by using the Carvalho et al. (2020) and Dors et al. (2019) calibrations (see Eq. 12 and 7) respectively, while $Z_{C43}$ through the semi-empirical calibration (Eq. 6) by Dors et al. (2019). The linear fittings (coefficients not shown) and their corresponding Pearson Correlation Coefficient values (R) are shown in Fig. 7. It was found that the R value indicates a weak correlation between $ZR$ and $\log U$ via $O32$. However, when $\log U$ is derived through $C3C4$, the R values indicate a moderate correlation with $ZR$. These distinct results could be an indication that these emission line ratios ($O32$ and $C3C4$) trace the ionization degree of different gas regions in NLRs of AGNs, influencing the $Z$ estimates. This is because both $C43$ and $R_{23}$ calibrations take into account the ionization degree of the gas via $C3C4$ and $O32$, respectively, i.e. they are bi-parametric calibrations (see Zhu et al. 2024 and references therein).

The moderate correlation between $\log U$ via $C3C4$ and $ZR$ could be attributed, in part, to the $C^{3+}$ ion being concentrated in inner/denser layers with a higher ionization degree and, probably, with distinct abundances than outer regions containing most of the $O^+$ and $O^{2+}$ ions. Indeed, optical spatially resolved gas-phase metallicity estimates in local ($z \lesssim 0.013$) Seyfert galaxies by Armah et al. (2024) show an increase of O/H ($\sim 0.2$ dex) and a decrease of $N_e$ (by a factor of $\sim 3$) with the increase of the nebular radius for most of the objects analyzed (see also Revalski et al. 2018). Moreover, analysis of a large sample of 80 quasars at $z \sim 3$ by Guo et al. (2020) shows a radial decline of the flux of the C IV$\lambda$1549 and He II$\lambda$1640 emission lines, indicating a variation of abundances and/or ionization degree with the nebular radius. Thus, the $ZR$ discrepancy could be interpreted as a result of metallicities derived from ultraviolet emission lines (specifically the $C43$) are representative of inner gas regions of AGNs, whilst the $R_{23}$ of outer layers. If this supposition is correct, we could derive a correlation between $ZR=Z_{C43}/Z_{R_{23}}$ and the electron density. In Fig. 7, top panel, the logarithm of the electron density (in cm$^{-3}$) versus $ZR$ is shown. Although the linear fitting indicates a correlation, the Pearson Correlation Coefficient value (R) shows that it is weak. The electron density was derived from the [S II]$\lambda$6716/$\lambda$6731 line ratio, which traces the density in outer layers with low abundance of $C^{3+}$, i.e. the $N_e$ values shown in Fig. 7 are not representative for the gas region where most of this ion is concentrated. Electron density estimates via lines emitted by ions with higher ionization potential values than for S$^+$ (e.g. [Ar IV]$\lambda$4711/$\lambda$4740; see Congiu et al. 2017; Cerqueira-Campos et al. 2021; Holden et al. 2023; Binette et al. 2024) are need to improve the above analysis. In any case, the assumption that $C3C4$ traces the ionization degree of inner layers than the $O32$ is supported by the lower $U$ values derived from the latter. To show that, in Fig. 8, the $\log U$ values derived from $C3C4$ are compared with those via $O32$ for our sample (listed in Table 4), where the calibrations by Zhu et al. (2024) are also included. It can be seen that for most of the cases and independently of the calibrations assumed, $C3C4$ yields higher $U$ values than $O32$.

As an additional test to verify if the $C43$ and $R_{23}$ indexes trace the metallicity of distinct gas regions in AGNs, we build photoionization models to analyze the ionization structure of the carbon and oxygen. We assume as nebular parameters a spherical geometry, solar metallicity, $\alpha_{ox} = -1.1$, and three values for $\log U$: $-1.0, -2.5$ and $-3.5$ dex. These parameter values are typical of NLRs (e.g. Pérez-Montero et al. 2019). In Fig. 9, the resulting ionization structure for the carbon ($C^+$, $C^{2+}$, $C^{3+}$), oxygen ($O^+$, $O^{2+}$) and hydrogen ($H^0$, $H^+$) are shown. We can note that:





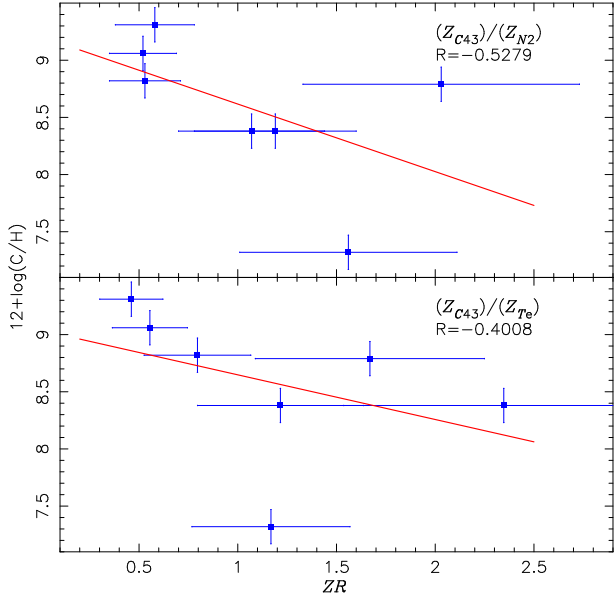

**Figure 10.** Carbon abundance [in units of 12+log(C/H)] versus the ratio ($ZR$) between the metallicity derived through the $C43$ index and $T_e$-method (lower panel) and $N2$ index (upper panel). Points represent the estimates for our AGN sample, while the lines are the linear fittings (coefficients not shown) to them.

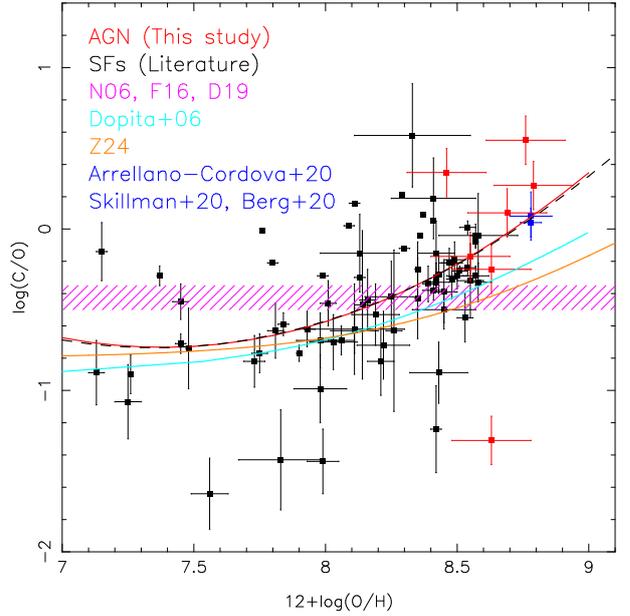

**Figure 11.** Logarithm of the C/O abundance ratio versus the oxygen abundance [in units of 12+log(O/H)]. Red points represent the estimates for our AGN sample (see Table 4). Black points represent SF estimates derived by Walter et al. (1992), Garnett et al. (1999), Esteban et al. (2009, 2014), Berg et al. (2016), Peña-Guerrero et al. (2017), Skillman et al. (2020), and Arellano-Córdova et al. (2020). Blue points represent the extrapolations to the nuclear region (galactocentric distance equal to 0) of the radial abundance gradients derived for the Milky Way by Arellano-Córdova et al. (2020) and for M 101 by Skillman et al. (2020) and Berg et al. (2020). The cyan and orange curves represent the abundance relations assumed in photoionization models by Dopita et al. (2006) and Zhu et al. (2024), respectively. The dashed area represents the range $-0.5 \leqq \log(C/O) \leqq -0.35$ assumed in photoionization models by Nagao et al. (2006), Feltre et al. (2016) and Dors et al. (2019). The red curve shows the abundance relation obtained taking into account both AGNs and SFs (Eq. 30) and the dashe line is the same but only considering SFs.

- Independently of the value assumed for log $U$, $C^{2+}$ and $O^{2+}$ occupy similar gas regions. This result shows the reliability of C/O abundance estimates for AGNs via the C III]/O III] line ratio.

- $C^{3+}$ and $O^+$ are concentrated preferably in inner and outer layers, respectively, independently of the log $U$ assumed. Thus, for AGNs with radial abundance profiles, naturally, $C43$ and $R_{23}$ indexes tend to represent the metallicity of distinct gas regions. In any case, any calibration must be applied by assuming emission line ratios measured from integrated spectra, in which the influence of radial abundance profiles on $Z$ estimates is minimized. Thus, the discrepancy $ZR$ observed in the present study, probably, is not due to the $C^{3+}$ and $O^+$ ions being concentrated in distinct gas regions.

- $C^+$ extends into the neutral hydrogen zone, being its fraction in the ionized region about null for log $U = -1.0$. This result makes it incorrect to use $C^+$ in the C/H abundance estimates.

Similar results are also derived for H II regions, as shown by Berg et al. (2019), who built photoionization models simulating H II regions to reproduce the $C^{3+}/C^{2+}$ ionic abundance ratio. In their models, it was necessary to require high ionization parameters (log $U \sim -1.5$), being the typical values for H II regions derived from optical lines in order of log $U \sim -2.5$ (e.g. Dors et al. 2011; Zinchenko et al. 2019; Ji & Yan 2022; Grasha et al. 2022; Garner et al. 2024). The fact that $C^+$ extends into the neutral zone, in principle, precludes its use in determining the C/H total abundance, as for $O^0$ in the O/H abundance calculations.

Concerning the carbon influence on discrepancies of the $Z$ estimations via UV and optical lines, in Fig. 10, values of the carbon abundance [in units of 12+log(C/H)] for our sample of objects versus the ratio ($ZR$) between metallicity estimates derived through the $C43$ calibration and the $T_e$-method (lower panel) and the $N2$ calibration (upper panel) are shown. This plot also presents a linear fitting (coefficients not shown) to the points and their corresponding R values. Despite the R values indicating a weak correlation, we can see that $ZR$ increases when the carbon abundance decreases. Since

the C/O-O/H abundance relation assumed as the input parameter in photoionization models deeply influences predictions of carbon emission lines, the discrepancy found above can result from incorrect suppositions of this relation in photoionization models. Distinct carbon-oxygen relations have been adopted in metallicity studies of AGNs. Nakajima et al. (2018) built a large grid of photoionization models to derive ultraviolet diagnostic diagrams used to classify and obtain nebular parameters of galaxies. These authors investigated models with distinct C/O-O/H relations and found that an enhanced C/O abundance ratio (up to the solar value) is needed for models with metallicities $(Z/Z_\odot) \sim 0.2$ to reproduce UV emission-line ratios of galaxies at $z = 2 - 4$. The photoionization models used by Dors et al. (2019) to derive the $C43$-$Z$ relation assume a fixed $\log(C/O)_\odot = -0.50$, a similar value assumed in the NLR models of Nagao et al. (2006) and Feltre et al. (2016). Finally, Zhu et al. (2024), opposite to the majority of previous studies that use nebular estimates, assumed in their photoionization models a C/O-O/H abundance relation derived from stellar abundance data in the Milky Way provided by Nicholls et al. (2017).

Our direct carbon abundance estimates can be used to constraint a more representative (C/O)-$Z$ relation for NLRs of AGNs. However, due to our small sample of objects for which it was possible to estimate C/H (7 objects) and the narrow range of metallicity





of them [0.6 ≲ (Z/Z$_\odot$) ≲ 1.3[5]], it is desirable to combine our estimates with those for SFs. This practice was adopted in other previous studies dedicated to derive abundance relations of other elements in NLRs of AGNs (e.g. Pérez-Montero et al. 2019, 2023; Monteiro & Dors 2021; Armah et al. 2021; Peluso et al. 2023; Dors et al. 2022, 2023, 2024b) and low-ionization nuclear emission-line region galaxies (LINERs, e.g. Oliveira et al. 2024). Moreover, comparing the relative abundance (e.g. N/O, C/O) of objects of distinct classes yields important information on the stellar evolution and chemical ISM enrichment in different environments. In this regard, in Fig. 11, a plot of log(C/O) versus 12+log(O/H), our abundance estimates (red points) are plotted together with those for SFs (black points). A dashed area representing the fixed log(C/O) values assumed in previous calibrations is also shown in this figure. The SF abundance estimates, taken from the literature obtained by distinct authors and for local ($z \lesssim 0.2$) objects, are derived either from UV C III]$\lambda$1909/O III]$\lambda$1666 collisional excited lines or from optical O II and C II recombination lines, both combined with the [O III]$\lambda$5007 (e.g. Esteban et al. 2009; Berg et al. 2016). These distinct methodologies to derive abundance can contribute to the scatter of C/O abundances at fixed O/H values. In fact, even in the best case, where the C II$\lambda$4267 and O II$\lambda$4649 recombination lines are measured, the O$^{2+}$ ionic abundance is always derived from optically excited lines (Skillman et al. 2020). In Fig. 11 are also included:

- The (C/O)-Z relation derived for H II regions by Dopita et al. (2006), which is based on results of local SFs compiled by Garnett et al. (1999):

$$(C/H) = 6.0 \times 10^{-5} \times (Z/Z_\odot) + 2.0 \times 10^{-4} \times (Z/Z_\odot)^2. \quad (29)$$

- The Eq. 13 assumed by Zhu et al. (2024) and derived from stellar abundance estimates (Nicholls et al. 2017).
- Values of the nuclear interception (radial distance equal to 0) of [12+log(O/H); log(C/O)] radial gradients derived for the Milky Way (MW) by Arellano-Córdova et al. (2020) and for M 101 by Skillman et al. (2020) and Berg et al. (2020). These values correspond to [8.78±0.08, 0.05±0.15]$_{MW}$ and [8.78±0.04, 0.158±0.09]$_{M101}$ and are represented by blue points.

We can note in Fig. 11 that AGNs present similar C/O abundances than the most metallic SFs (mainly composed by inner disc H II regions) and they are in agreement with those from the extrapolation of the radial gradients in the Milky Way and M 101. Also, our small sample of AGNs seem to indicate that, similarly to the (N/O)-(O/H) relation (Dors et al. 2024b), this object class follows the (C/O)-O/H relation for SFs. Only an AGN, i.e. NGC 4507, presents a very low C/O abundance, deviating from the trend shown by AGNs and SFs with similar O/H abundances, although some few SFs are showing a similar deviation. A fitting considering all the points (AGNs+SFs) results in

$$\log(C/O) = 0.41(\pm 0.24) x^2 - 6.07(\pm 3.87) x + 21.69(\pm 15.45), \quad (30)$$

where x=12+log(O/H). This relation is represented in Fig. 11 by a red curve. We also produced a fitting considering only SF estimates and found the same relation (represented in Fig. 11 by a dashed black line) than that assuming AGNs+SFs. It is worth noting that our relation results in:

- Values of log(C/O) higher by ∼ 0.2 dex than those from the Dopita et al. (2006) and Zhu et al. (2024) relations, while this

---

[5] Metallicity range derived by assuming the $T_e$-method (see Table 4).

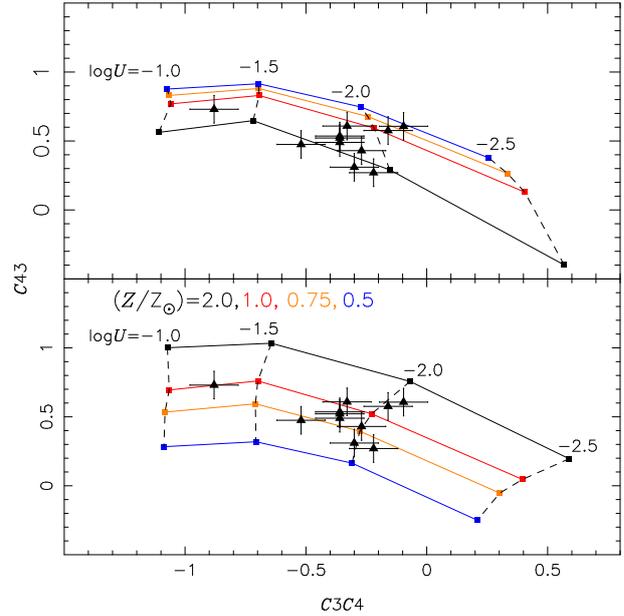

**Figure 12.** $C43=\log[(\text{C IV}]\lambda1549+\text{C III}]\lambda1909)/\text{He II}\lambda1640]$ versus $C3C4=\log[\text{C III}]\lambda1909/\text{C IV}]\lambda1549]$. Lower Panel: The lines connect the results of photoionization models built assuming the (C/O)-(O/H) abundance ratio given by Eq. 30. Solid and dashed lines connect model results with the same metallicity and log $U$, respectively, as indicated. The triangles represent our observational sample. Upper panel: As the lower panel but for photoionization models assuming a fixed value of log(C/O)$_\odot = -0.30$ (see Sect. 3.3).

difference reaches up to ∼ 0.4 dex in the region occupied by the AGNs.
- Similar abundances to those derived from nuclear extrapolations of radial abundance gradients in the Milky Way and M 101.
- The fixed values of log(C/O) ∼ −0.4, assumed in the photoionization models by Nagao et al. (2006), Feltre et al. (2016) and Dors et al. (2019), do not seem to be representative for NLRs.

To test if the use of our (C/O)-(O/H) abundance relation alleviates the $ZR$ discrepancy derived when metallicities from UV lines are compared with those via optical lines (see Fig. 5), we build a new photoionization model grid by assuming as input parameter the (C/O)-(O/H) relation represented by Eq. 30. After comparing model results with our observational data, we derive new values of $Z_{C43}$ for each object in our sample from interpolation between the models. This methodology is considered to derive semi-empirical calibrations (e.g. Castro et al. 2017). In a subsequent paper (Dors et al. in preparation), we will produce a new $C43$-$Z$ calibration by assuming abundances from the Eq. 30, a wide range of nebular parameters (e.g. $\alpha_{ox}$, $N_e$) and a large sample of observational data. The present study is limited to only analyze the influence of nebular parameters on $Z$ estimates via the $C43$ index calibration.

In the lower panel of Fig. 12, a $C43$ versus $C3C4$ diagram, our observational data are compared with the results of our new photoionization model grid. In the bottom part of Fig. 13, we compare the $Z$ values obtained from Fig. 12, lower panel, with those derived using the $T_e$-method (listed in Table 4) for the objects in our sample for which this comparison is possible. It can be seen a good agreement between the estimates, being NGC 5506 the only object presenting a large disagreement between the estimates. In the top part of Fig. 13, the metallicity ratio $ZR$ versus the metallicity estimates via the $T_e$-method is shown. The black line represents a linear fitting (coefficients not shown) to all points. As found when the $C43$ calibrations were considered (see Fig. 5), a correlation between $ZR$ and $Z_{T_e}$ remains (R ∼ −0.6), with the mean value for the metallicity ratio equal to $<ZR> = 1.22 \pm 0.69$. However, excluding estimates for NGC 5506, a new linear fitting (red line) indicates a one-by-one metallicity correlation, with R ∼ 0 and $<ZR> = 0.96 \pm 0.16$.

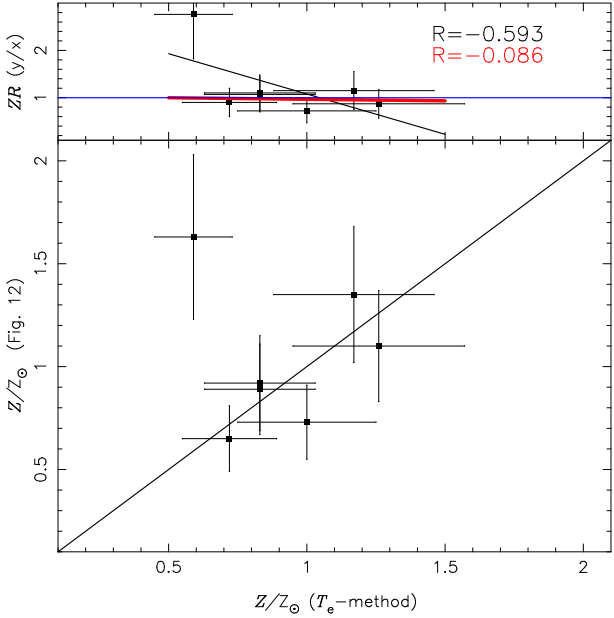
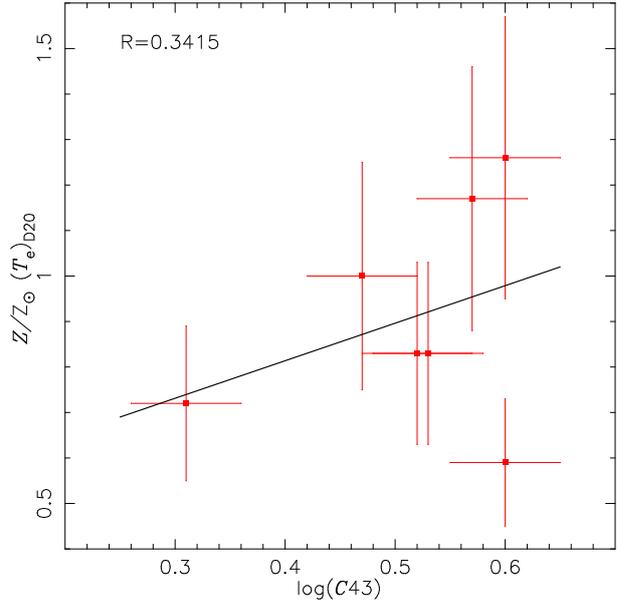

**Figure 13.** As Fig. 5 (left panel) but for metallicities (in relation to the solar one) derived from the interpolation of the models in the lower panel of Fig 12 versus those via the $T_e$-method. The black line represents a linear fitting to all points, whose Pearson Correlation Coefficient (R) is indicated. The red line is like the black one but obtained excluding the outlier NGC 5576 (see text). The resulting R-value of this fitting is indicated in red.

**Figure 14.** Metallicity ($Z$), relative to the solar value, calculated through the $T_e$-method using O/H abundances, is plotted against the observational $C43$ values. Red points represent the values for seven Seyfert 2 galaxies ($z < 0.03$) from our sample (see Table 4) for which it was possible to apply the $T_e$-method. The line represents a fit to the points (coefficients not shown). The Pearson correlation coefficient value (R) is also shown.

We emphasize that although there is a clear need for a larger number of metallicity estimates via UV lines and the $T_e$-method to improve the $Z$ comparison, the result in Fig. 13 indicates an agreement between UV estimates and those via the $T_e$-method. Thus, we argue that the relation between carbon and oxygen abundances is the main source of discrepancy between metallicities based on carbon ultraviolet lines and those derived by direct estimates adopting optical emission lines.

### 3.3 Influence of (C/O)-$Z$ relations

Since some previous UV calibrations (e.g. Dors et al. 2019) have assumed fixed C/O values in their photoionization models, it is helpful to analyze how $Z$ and $U$ derived in this case differ from those obtained through models that assume a more realistic abundance relation. For this purpose, we constructed a photoionization model grid assuming a fixed value of $\log(C/O)_\odot = -0.30$. The results of these models are compared with the observational data in the upper panel of Fig. 12. It can be seen a clear difference with the model results (lower panel of Fig. 12) that assume the (C/O)–(O/H) relation represented by Eq. 30. The assumption that C/O does not vary with metallicity implies that objects with high $C43$ values exhibit the lowest $Z$ values, i.e., $C43 \propto 1/Z$. In contrast, the models assuming the relation represented by Eq. 30 indicate that objects with high $C43$ values exhibit the highest $Z$ values, i.e., $C43 \propto Z$. The result that $C43$ must increase with $Z$ is confirmed by our direct estimates shown in Fig. 14, where $Z$ values derived through the $T_e$-method via O/H abundances are plotted against the observational values of $C43$ for our AGN sample. Despite the large scatter of the points, the R value indicates a positive correlation between $Z$ and $C43$, represented by the red line (a linear fit to the points).

The discrepancy above has a deep impact on the $Z$ estimates. To show that, in Fig. 15, a comparison between the $Z$ and $U$ values derived from both grids of models presented in Fig. 12 is shown. It can be noted: (*i*) the use of photoionization models assuming the (C/O)–(O/H) relation represented by Eq. 30 result in higher $Z$ values than those assuming a fixed C/O solar value for the high metallicity regime and vice versa, and (*ii*) lower (∼ 0.3 dex) $\log U$ values are derived by models assuming the Eq. 30 instead of a fixed solar C/O value.

## 4 CONCLUSIONS

Metallicity ($Z$) estimates of the gas phase of high-$z$ galaxies are mainly derived by calibrations between strong UV emission lines and this nebular parameter. These estimates $Z$ have shown to be discrepant in relation to those via optical emission lines, being the origin of this discrepancy ($ZR$) poorly defined. To investigate the origin of $ZR$ in NLRs of AGNs, we compiled from the literature spectroscopic observational data in the UV (1000 < $\lambda$(Å) < 2000) and optical (3000 < $\lambda$(Å) < 7000) wavelength ranges of a sample of 11 objects (9 at $z < 0.4$ and 2 at $z \sim 2.4$). We estimated the metallicity for each object of this sample adopting a semi-empirical calibration assuming the $C43 = \log[(C\,\textsc{iv}\lambda 1549 + C\,\textsc{iii}]\lambda 1909)/\textrm{He}\,\textsc{ii}\lambda 1640]$ index as a metallicity tracer. These estimates were compared with those derived through direct electron temperature measurements, i.e. the $T_e$-method. Also, calibrations based on optical strong lines were considered when it was not possible to apply the $T_e$-method (4 out 11 objects). The source of the discrepancy was investigated in terms of the ionization parameter ($U$), electron density ($N_e$), and carbon abundance (C/H) derived through the $T_e$-method. A comparison of the ionization parameter estimates based on UV and optical emission lines was also carried out. We found the following results:

- $C43$ index tends to overestimate the metallicity for $(Z/Z_{\odot}) \lesssim 1.0$ and underestimate it for higher $Z$ values in comparison to optical estimates.
- Ionization parameter values derived from the $C3C4$=C III]$\lambda$1909/C IV$\lambda$1549 line ratio tend to be higher than those derived from $O32$=[O III]$\lambda$5007/[O II]$\lambda$3727.
- Photoionization model simulation of the ionization structure of AGNs indicates that the $C^{3+}$ ion is emitted mainly in inner parts of the nebulae, while $O^+$ in outer parts. This result indicates that $C3C4$ and $O32$ measure the ionization degree of distinct gas regions in AGNs, possibly having some influence on $Z$ estimates via calibrations adopting these line ratios to mitigate the effect of ionization parameter in $Z$ estimates.
- We found a weak correlation of the ratio ($ZR$) between metallicities via $C43$ and optical lines with the ionization parameter and electron density. Otherwise, a moderate correlation was found between $ZR$ and direct estimates of C/H, implying that the (C/O)-$Z$ relations previously adopted in photoionization models for UV carbon-line calibrations might not be accurate for AGNs.
- We derived a new (C/O)-$Z$ relation by combining direct abundance estimates from our sample with abundances from a large number of local star-forming objects. Photoionization models based on this new relation yield $Z$ values comparable to those calculated using the $T_e$-method.

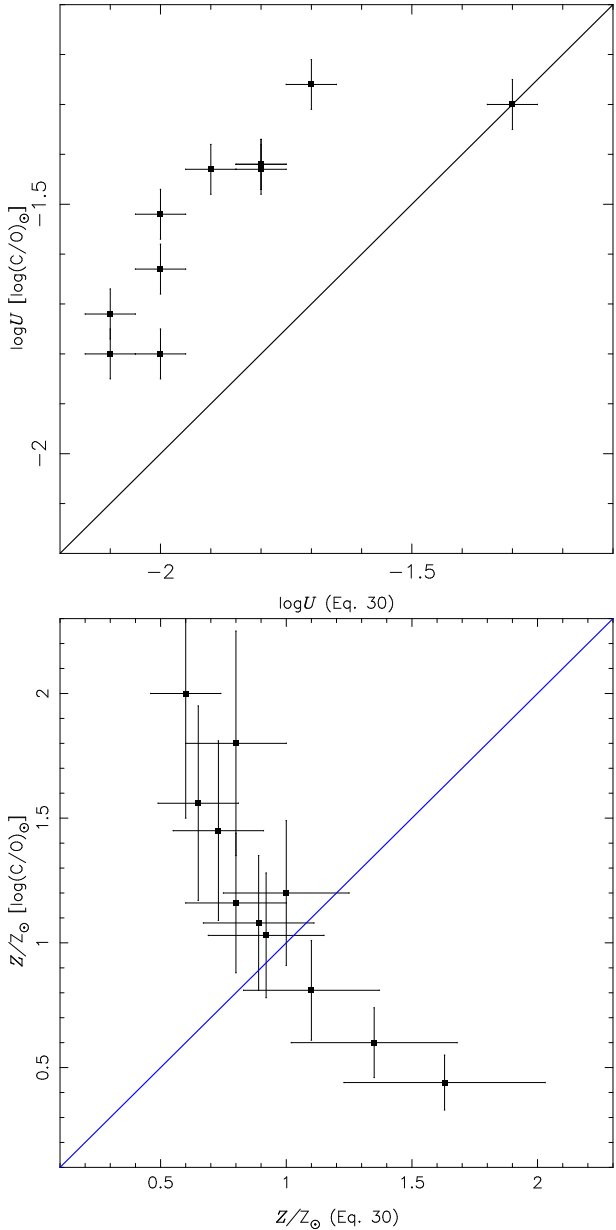

**Figure 15.** Lower panel: Comparison between the metallicity for our AGN sample derived by interpolating the photoionization models (shown in the lower panel of Fig. 12) assuming the (C/O)-(O/H) relation represented by Eq. 30 versus those from models (shown in the lower panel of Fig. 12) assuming a fixed $\log(C/O)_{\odot} = -0.30$. The line indicates the equality between the estimates. Upper panel: As lower panel but for the logarithm of the ionization parameter ($U$).

## ACKNOWLEDGEMENTS

OLD is grateful to Fundação de Amparo à Pesquisa do Estado de São Paulo (FAPESP), process number 2022/07066-6, and to Conselho Nacional de Desenvolvimento Científico e Tecnológico (CNPq). RAR acknowledges the support from Conselho Nacional de Desenvolvimento Científico e Tecnológico (CNPq; Proj. 303450/2022-3, 403398/2023-1, & 441722/2023-7), Fundação de Amparo à pesquisa do Estado do Rio Grande do Sul (FAPERGS; Proj. 21/2551-0002018-0), and Coordenação de Aperfeiçoamento de Pessoal de Nível Superior (CAPES; Proj. 88887.894973/2023-00). RR acknowledges support from Conselho Nacional de Desenvolvimento Científico e Tecnológico ( CNPq, Proj. 311223/2020-6, 304927/2017-1, 400352/2016-8, and 404238/2021-1), Fundação de amparo à pesquisa do Rio Grande do Sul (FAPERGS, Proj. 19/1750-2 and 24/2551-0001282-6) and Coordenação de Aperfeiçoamento de Pessoal de Nível Superior (CAPES, Proj. 0001).

## DATA AVAILABILITY

The data underlying this article will be shared on reasonable request to the corresponding author.

This paper has been typeset from a T$_{\rm E}$X/L$^{\rm A}$T$_{\rm E}$X file prepared by the author.